\documentclass[useAMS,usenatbib]{mn2e}

\usepackage{amsmath}
\usepackage{graphicx}
\usepackage{float}
\usepackage{subcaption}
\usepackage{gensymb}
\usepackage{amssymb}
\usepackage{tabularx}
\usepackage{subfig}
\usepackage{lineno}
\usepackage{url}
\usepackage{float}
\DeclareRobustCommand{\ion}[2]{%
\relax\ifmmode
\ifx\testbx\f@series
{\mathbf{#1\,\mathsc{#2}}}\else
{\mathrm{#1\,\mathsc{#2}}}\fi
\else\textup{#1\,{\mdseries\textsc{#2}}}%
\fi}

\newcommand{\SFNUM}{$201 \:$} 

\title[Spatially Resolved Environment Quenching]{The SAMI Galaxy Survey: Spatially resolving the environmental quenching of star formation in GAMA galaxies}
\author[Schaefer et al.]
{\parbox{\textwidth}
{A.~L.~Schaefer$^{1,2,3}$\thanks{E-mail:  \texttt{schaefer@physics.usyd.edu.au}},
S.~M.~Croom$^{1,3}$, 
J.~T.~Allen$^{1,3}$,
S.~Brough$^{2,3}$,
A.~M.~Medling$^{4}$,
I.-T.~Ho$^{4,5}$,
N.~Scott$^{1,3}$,
S.~N.~Richards$^{1,2,3}$,
M.~B.~Pracy$^{1}$,
M.~L.~P.~Gunawardhana$^{6}$,
P.~Norberg$^{6}$,
M.~Alpaslan$^{7}$,
A.~E.~Bauer$^{2}$,
K.~Bekki$^{9}$
J.~Bland-Hawthorn$^{1}$,
J.~V.~Bloom$^{1,3}$,
J.~J.~Bryant$^{1,2,3}$,
W.~J.~Couch$^{8}$,
S.~P.~Driver$^{9,10}$,
L.~M.~R.~Fogarty$^{1}$,
C.~Foster$^{2}$,
G.~Goldstein$^{12}$
A.~W.~Green$^{2}$,
A.~M.~Hopkins$^{2}$,
I.~S.~Konstantopoulos$^{2,11}$,
J.~S.~Lawrence$^{2}$,
A.~R.~L\'opez-S\'anchez$^{2,12}$
N.~P.~F.~Lorente$^{2}$,
M.~S.~Owers$^{2,12}$,
R.~Sharp$^{4}$
S.~M.~Sweet$^{4}$,
E.~N.~Taylor$^{8}$,
J.~van~de~Sande$^{1}$,
C.~J.~Walcher$^{13}$
O.~I.~Wong$^{3,9}$
}\vspace{0.4cm} \\
\parbox{\textwidth}{$^{1}$Sydney Institute for Astronomy, School of Physics, University of Sydney, NSW 2006, Australia\\
$^{2}$Australian Astronomical Observatory (AAO), PO Box 915, North Ryde, NSW 1670, Australia\\
$^{3}$CAASTRO: ARC Centre of Excellence for All-sky Astrophysics\\
$^{4}$Research School of Astronomy and Astrophysics, Australian National University, Canberra, ACT 2611, Australia\\
$^{5}$Institute for Astronomy, University of Hawaii, 2680 Woodlawn Drive, Honolulu, HI 96822 USA\\
$^{6}$Institute for Computation Cosmology and Centre for Extragalactic Astronomy, Department of Physics, Durham University, Durham, DH1 3LE, UK\\
$^{7}$NASA Ames Research Center, N232, Moffett Field, Mountain View, CA 94035, United States\\
$^{8}$Centre for Astrophysics and Supercomputing, Swinburne University of Technology, PO Box 218, Hawthorn, VIC 3122, Australia\\
$^{9}$International Centre for Radio Astronomy Research, University of Western Australia, 35 Stirling Highway, Crawley, WA, 6009, Australia\\
$^{10}$SUPA School of Physics \& Astronomy, University of St Andrews KY16 9SS Scotland\\
$^{11}$Envizi, Suite 213, National Innovation Centre, Australian Technology Park, 4 Cornwallis Street, Eveleigh NSW 2015, Australia\\
$^{12}$Department of Physics and Astronomy, Macquarie University, NSW 2109, Australia \\
$^{13}$ Leibniz-Institut für Astrophysik Potsdam (AIP) An der Sternwarte 16 D-14482, Potsdam, Germany}
}

\begin{document}

\date{2016}

\pagerange{\pageref{firstpage}--\pageref{lastpage}} \pubyear{2016}

\maketitle

\label{firstpage}

\begin{abstract}
We use data from the Sydney-AAO Multi-Object Integral Field Spectrograph (SAMI) Galaxy Survey and the Galaxy And Mass Assembly (GAMA) survey to investigate the spatially-resolved signatures of the environmental quenching of star formation in galaxies. Using dust-corrected measurements of the distribution of H$\alpha$ emission we measure the radial profiles of star formation in a sample of \SFNUM star-forming galaxies covering three orders of magnitude in stellar mass ($\rm{M}_{*}$; $10^{8.1}$-$10^{10.95} \, \mathrm{M}_{\odot}$) and in $5^{th}$ nearest neighbour local environment density ($\Sigma_{5}$; $10^{-1.3}$- $10^{2.1} \, \mathrm{Mpc}^{-2}$). We show that star formation rate gradients in galaxies are steeper in dense ($\log_{10}(\Sigma_{5}/\mathrm{Mpc^{2}})>0.5$) environments by $0.58 \pm 0.29 \,  \mathrm{dex} \, \mathrm{r_{e}}^{-1}$ in galaxies with stellar masses in the range $10^{10} < \mathrm{M_{*}}/\mathrm{M_{\odot}} < 10^{11}$ and that this steepening is accompanied by a reduction in the integrated star formation rate. However, for any given stellar mass or environment density the star-formation morphology of galaxies shows large scatter. We also measure the degree to which the star formation is centrally concentrated using the unitless scale-radius ratio ($r_{50,H\alpha}/r_{50,cont}$), which compares the extent of ongoing star formation to previous star formation. With this metric we find that the fraction of galaxies with centrally concentrated star formation increases with environment density, from $\sim 5 \pm 4 \%$ in low-density environments ($\log_{10}(\Sigma_{5}/\mathrm{Mpc^{2}})<0.0$) to $30\pm 15 \%$ in the highest density environments ($\log_{10}(\Sigma_{5}/\mathrm{Mpc^{2}})>1.0$). These lines of evidence strongly suggest that with increasing local environment density the star formation in galaxies is suppressed, and that this starts in their outskirts such that quenching occurs in an outside-in fashion in dense environments and is not instantaneous.

\end{abstract}


\section{Introduction}
The process of star formation is critical to the evolution of galaxies. The rate of star formation past and present has a significant effect on the optical colours and morphology of a given galaxy \citep{DG92}. It has become apparent that the environment within which a galaxy is situated plays an important role in that galaxy's development \citep[e.g.][]{HubbleAndHumason31,Oemler74,Dressler80}. The presence of a relationship between galaxy environment and star-forming properties suggests that the transformation from star-forming to quiescent, a process called quenching, could be affected by the environment.

A number of mechanisms have been proposed that could cause quenching to occur. These mechanisms generally involve the removal of the gas supply that fuels star formation. 

Ram pressure stripping has been identified as a potential method for reducing the amount of available gas in a galaxy disc \citep{Gunn72} and in the surrounding halo \citep{McCarthy08}. The role of ram pressure stripping in quenching star formation in cluster galaxies has been well established. The best evidence for ram pressure stripping acting to quench star formation within clusters comes from unresolved measurements of neutral hydrogen in cluster galaxies \citep[e.g.][]{Giovanelli&Haynes85,Solanes2001,Cortese2011}. The signatures of ram pressure stripping include the confinement of star formation to the central regions of the galaxy \citep{Koopmann04b,Cortese2012}, and the presence of a tail visible in H$\alpha$, neutral hydrogen, or both \citep{Balsara1994}. \cite{Boselli2006} utilised multiwavelength imaging to constrain the stellar population gradients within a galaxy to confirm that ram pressure stripping acts on late type galaxies in clusters. This process is also capable of acting in compact groups \citep{Rasmussen08} as well as in less concentrated environments \citep{Bekki2009,Merluzzi13}. \cite{Nichols11} successfully explain the gas fractions of dwarfs in the Milky Way $+$ M31 system by modelling gas removal by ram pressure from these objects analytically. However, \cite{Rasmussen08} find that ram pressure stripping alone cannot explain the gas deficiencies in more massive group galaxies, with masses of approximately $4\times 10^{10} \, \rm{M}_{\odot}$ or greater.

It has been pointed out that other transport processes could be responsible for the removal of gas from galaxy discs in dense environments. In particular, turbulent viscous and inviscid stripping have been suggested to be significant mechanisms for gas removal in dense environments \citep{Nulsen1982,Roediger13}. When the velocity of the galaxy through the intergalactic medium is subsonic, viscous and turbulent mixing between the two media can become important in liberating gas from galaxies. The timescale for removing the gas from galaxies undergoing viscous stripping is shorter than for ram-pressure stripping \citep{Nulsen1982}. Modelling has shown that the morphological signatures of these mechanisms may be similar to those of ram-pressure stripping \citep[e.g.][]{BoselliAndGavazzi2006,Roediger14}.

The timescale on which a disc galaxy would deplete its gas through ordinary star formation is on the order of a few Gyr \citep{Miller79, Tinsley80}. In order to explain the existence of gas-rich disc galaxies in the Universe today, it is necessary to suppose that the gas within their discs has been replenished by infall from the intergalactic medium \citep[e.g.][]{Larson72}. If this supply is cut off, then quenching will occur in a process called strangulation. This is most likely to transpire when the gas envelope surrounding spiral galaxies is swept away as the galaxy and its halo enter a cluster or group and fall through the denser intergalactic medium of these environments \citep{Larson80}. Strangulation is likely to quench star formation over a period of several Gyrs \citep{McCarthy08} once the gas in the disc ceases to be replenished. \cite{Peng15} estimate that strangulation is responsible for quenching approximately 50\% of the passive galaxy population today, though they do not investigate the dependence of this fraction on environment density. Strangulation is predicted to occur in galaxy groups by \cite{Kawata2008}. Some simulations \citep[e.g.][]{Bekki02} suggest that the formation of anaemic spirals, systems with uniformly suppressed star formation \citep{vandenbergh91,Elmegreen02}, can be explained by strangulation. This process is not expected to produce the same spatial distribution of star formation as other processes such as ram-pressure stripping while quenching is taking place.


Galaxies in high-density environments such as clusters or galaxy groups may also experience tidal interactions with their nearest neighbours, or the group or cluster gravitational potential. These tidal interactions often have the effect of driving gas towards the centre of the galaxy and triggering circumnuclear starbursts \citep[e.g.][]{Heckman90}. These starburst episodes can deplete the gas reservoir of the galaxy, causing it to become quenched. Simulations \citep[e.g.][]{Hernquist89,Moreno2015} have shown that gravitational instabilities driven by tidal interactions will drive a large fraction of the gas in a galaxy towards the centre and enhance star formation on short timescales. This will have the effect of producing galaxies with centrally-concentrated star formation shortly after an interaction.

\cite{Faber07} argued that gas-rich major mergers between blue-sequence galaxies can induce a period of starburst, which rapidly depletes the interstellar gas within these systems. Following this merger, the remnant moves to the red sequence. However, \cite{Blanton06} notes that between $z=1$ and $z=0$ the number of blue-sequence galaxies is reduced by less than 10\%. This implies that major-merger-driven quenching cannot have been the dominant mechanism for decreasing star formation in the second half of the Universe's history.

Recently, large-scale spectroscopic and photometric surveys have been able to make significant progress in understanding the relationship between galaxy environment and star formation. Modern multi-object spectroscopic surveys such as the 2 Degree Field Galaxy Redshift Survey \citep[2dFGRS;][]{Colless01}, the Sloan Digital Sky Survey \citep[SDSS;][]{York2000} and the Galaxy and Mass Assembly survey \citep[GAMA;][]{Driver11,Hopkins13} have allowed the determination of star formation rates in several hundred thousand galaxies. However, the gain in sample size afforded by these single-fibre spectroscopic surveys is offset by the fact that the star formation in each galaxy is reduced to an estimate of the integrated total, which can be affected by aperture bias. Consequently, this observational technique has been unable to identify galaxies that are in the process of quenching, and arguments involving the timing and frequency of quenching must be invoked to determine which mechanisms are producing the observed trends. 

A spectroscopic study of $521$ clusters in the SDSS by \cite{vonderLinden2010} indicated that the star formation rates of galaxies decline slowly during the infall into a cluster, with the most rapid quenching only occurring at the centres of clusters. The inferred quenching timescales were therefore long, roughly a few Gyrs, which is comparable to the cluster crossing time. This conclusion is in agreement with \cite{Lewis02} and \cite{Gomez03} who note the existence of galaxies with low specific star formation rates at large distances from the centres of clusters and conclude that rapid environmental quenching alone cannot explain the population of galaxies seen in the local Universe.

This picture appears to be inconsistent with the results of other work. Using the optical colours of galaxies, \cite{Balogh04} argue that once galaxy luminosity is controlled for, the dominant environmental trend is the changing ratio of red to blue galaxies, with very little change in the average colour of the blue galaxies. Similarly a spectroscopic investigation by \cite{Wijesinghe12} using GAMA data showed no correlation between the star formation rates of star-forming galaxies and the local environment density. In this sample, the trend was visible only when the passive galaxy population was included in the analysis, suggesting that it is the changing fraction of passive galaxies that is responsible for the observed environmental trends, and that quenching must therefore be either a rapid process or no longer proceeding in the local Universe. 

The ambiguity between fast and slow-mode quenching may in part arise from the different definitions of passive and star forming used by various teams \citep[see e.g.][]{Taylor15}, as well as the different methods of quantifying environment density employed. Moreover, these large-scale surveys are not able to investigate the spatial distribution of star formation in galaxies that are in the process of being quenched. 

The spatial properties of star formation in dense environments have been studied using narrow-band imaging of the H$\alpha$ distribution within galaxies \citep[e.g.][]{Koopmann04a, Koopmann04b,Gavazzi2006,Bretherton2013}. In the Virgo cluster, \cite{Koopmann04b} note that approximately half of $84$ observed spiral galaxies have spatially-truncated star formation while less than $10\%$ are anaemic (have globally reduced star formation). 
\cite{Welikala2008} and \cite{Welikala2009}, using spatially-resolved photometry from SDSS, suggested that the suppression of star formation in dense environments occurs predominantly in the centres of galaxies. \cite{Welikala2009} observed that the integrated star formation rates in star-forming galaxies decline with density, implying that the observed environmental trends cannot be completely explained by the morphology-density relation.
In lower density group and field environments, \cite{Brough13} used optical integral field spectroscopy to examine the radial distribution of star formation for $18$ $10^{10}\, \mathrm{M}_{\odot}$ galaxies and found no correlation between the star formation rate gradient and the local density. 

Several studies have also suggested that the stellar mass, bulge mass or other internal properties of a galaxy have a greater influence on whether it is quenched than does the environment \citep[e.g.][among others]{Peng2010,Bluck14,Pan15}. \cite{Peng2010} argued that quenching can be explained by two separable processes that depend on mass and environment. Mass quenching could be achieved by several mechanisms including AGN feedback, which either removes gas from the galaxy disc directly or prevents it accreting from the halo, or by feedback that is related to star-formation such as supernova winds.

While it is likely that all quenching mechanisms operate to some extent, it remains uncertain how dominant each mode is at a given environment density and galaxy mass. In this paper we investigate the radial distribution of star formation in galaxies observed as part of the Sydney-AAO Multi-object Integral Field Spectrograph (SAMI) Galaxy Survey \citep{Croom12, Allen15, Bryant15, Sharp15}. The application of spatially-resolved spectroscopy to this problem represents an important step towards a better understanding of the quenching processes in galaxies. With this technique applied to a large sample of galaxies, the spatial distribution of star formation in galaxies can be resolved and the direct results of the various quenching mechanisms can be observed. The broad range of stellar masses and environments targeted by SAMI make it an excellent survey for studying the spatial signatures of environmental quenching processes.

In Section~\ref{TargetSelection} we introduce the data used, our target selection and important ancillary data. Section~\ref{DataReduction} details the data reduction techniques employed by SAMI and the subsequent analysis of the flux-calibrated spectra. Results are presented in Section~\ref{Results} with a discussion and conclusion given in Sections~\ref{Discussion} and~\ref{Conclusion} respectively. Throughout this paper we assume a flat $\Lambda$CDM cosmology with $H_{0}=70$ km s$^{-1}$ Mpc$^{-1}$, $\Omega_{M}=0.27$ and $\Omega_{\Lambda}=0.73$ and adopt a \cite{Chabrier2003} stellar initial mass function.

\section{Data and Target Selection}\label{TargetSelection}

SAMI is a fibre-based integral field spectrograph capable of observing $12$ galaxies simultaneously \citep{Croom12}. Below we describe the full SAMI Galaxy Survey, which will include $\sim 3400$ objects.

\subsection{Target Selection}\label{target_selection}
SAMI Galaxy Survey targets were chosen from the parent GAMA survey. GAMA provides a high level of spectroscopic completeness \citep[$98.5\%$ in the regions from which SAMI targets were drawn;][]{Liske15}, providing spectroscopy of targets $2$ magnitudes deeper than the SDSS. Within the GAMA survey regions there is a large volume of complementary multi-wavelength data available, including radio continuum (\citealp[NVSS;][]{Condon98}, \citealp[FIRST;][]{Becker95}) and emission line data \citep[HIPASS;][]{Barnes01}, infrared \citep[UKIDSS;][]{Lawrence07} and ultraviolet \citep[\textit{GALEX};][]{Liske15} photometry, in addition to the SDSS and GAMA optical imaging, photometry and spectroscopy.
The SAMI targets were selected from a number of pseudo volume-limited samples, with each volume based on a spectroscopic redshift corrected for local flow effects \citep{Tonry2000} and galaxy stellar mass estimates of \cite{Taylor11}. Each volume is selected to be well above the sensitivity limits of the GAMA spectroscopic survey. SAMI will collect data for $\sim 3400$ galaxies in the redshift range $0.001<z<0.1$. 

The full SAMI survey also includes a complementary cluster sample of $\sim 600$ galaxies selected from eight clusters. High mass clusters are not well-represented in the GAMA survey and as such the SAMI cluster galaxies have been selected in a different way to the main sample. To ensure homogeneity for our sample, we do not include the cluster galaxies in this study. The target selection for the SAMI survey is discussed in detail by \cite{Bryant15}.

\subsection{SAMI Data}
The data used in the present study were obtained between $2013$ and $2015$ as part of the SAMI Galaxy Survey. Each of the $808$ galaxies comprising our input sample were observed with an offset dither pattern of approximately $7$ pointings to achieve uniform coverage, with each pointing being exposed for $1800\,\mathrm{s}$ and a total integration time of $12600 \,\mathrm{s}$ for each galaxy. 

The raw data were reduced with the 2dFDR pipeline \citep{Croom04}. This process resulted in row-stacked spectra that have been wavelength calibrated and subtracted of night sky continuum and emission line features. These spectra are combined into data cubes using a \textsc{python} pipeline\footnote{Astrophysics Source Code Library, \\ ascl:1407.006 \url{ascl.net/1407.006}} designed specifically for the construction of SAMI data cubes. The process of constructing the data cubes includes a correction for atmospheric dispersion, flux calibration and the removal of telluric absorption features. The data reduction is described in detail by \cite{Allen15} and \cite{Sharp15}. The data for each galaxy observed are divided between two cubes corresponding to the blue and red arms of the AAOmega spectrograph \citep{Sharp2006}. These spectral cubes cover wavelengths $\lambda \lambda 3700\mbox{--}5800$$\, \mathrm{\AA}$ with a spectral resolution of $\mathrm{R}=1810$ (at $\lambda=4800 \, \mathrm{\AA}$) for the blue and $\lambda \lambda 6300\mbox{--}7400 \, \mathrm{\AA}$ ~in the red at $\mathrm{R}=4260$ (at $\lambda=6850 \, \mathrm{\AA}$)\footnote{These spectral resolutions differ from the values quoted in previous SAMI papers. The latest resolution values are derived empirically from CuAr arc spectra by Van de Sande et al. \emph{in prep.}}. The data from each of the dithered pointings were regridded onto a $50\times 50$ array of $0\farcs 5$ square spatial picture elements (spaxels), each containing the spectrum of the target galaxy at that point. For a full discussion of the cubing process, see \cite{Sharp15}.

\subsection{Environment Density}\label{Env_density}
From an observational perspective it is difficult to define a single metric which fully describes the local environment of a galaxy. The metric used must be guided by the available data and the environmental processes of interest. For example, the effects of the gaseous intracluster medium on galaxy evolution are best quantified by using the X-ray properties of the cluster \citep[see e.g.][]{Owers12}, while galaxy merger or interaction rates can be studied using analysis of galaxy close pairs \citep[e.g.][]{Patton2000,Robotham14}. It is not clear what physical mechanism causes the environmental quenching of star formation, so there is no obvious choice as to which metric is appropriate to identify the various processes, although it is appropriate to select a measurement that is sensitive over a variety of density ranges. \cite{Muldrew12} constructed mock observational catalogues from cosmological simulations to examine the relationship between various environment density metrics and the underlying dark-matter distribution. They found that $n^{th}$ nearest neighbour estimators, defined as $\Sigma_{n}=\frac{n}{\pi \times r_{n}^{2}}$, where $r_{n}$ is the projected distance to the galaxy's $n^{th}$ nearest neighbour above some absolute magnitude limit, performed well at recovering the local density of dark matter. 

We use the fifth-nearest-neighbour local surface density measurement, $\Sigma_{5}$, to quantify the local environment around galaxies that have been targeted by SAMI. The high level of spectroscopic completeness of the GAMA survey means that it is well suited to calculating $\Sigma_{5}$. These environment measurements are performed on a density defining pseudo-volume-limited population of galaxies that have been observed spectroscopically by GAMA. This population includes all galaxies in the GAMA-II catalogue with reliable redshifts and K-corrected SDSS $r$-band absolute magnitudes, $M_{r}(z_{ref}=0,Q=1.03)<-18.5$, where $Q=1.03$ models the expected redshift evolution of $M_{r}$ \citep{Loveday15}. Only objects within $1000$ km s$^{-1}$ of a target galaxy contribute to the estimate of its local surface density and the observed surface density is scaled by the reciprocal of the survey completeness in that vicinity\footnote{The GAMA spectroscopic completeness, defined as the ratio of the number of objects with measured redshifts to the number of potential spectroscopic targets in a survey region, is extremely high. For the $808$ input SAMI galaxies the mean GAMA completeness in the surrounding region is $0.976$ with a standard deviation of $0.038$. As such, the reciprocal weighting of the $\Sigma_{5}$ measurements will not bias the results presented here.} \citep{Brough13}.

In practice the measurement of $\Sigma_{5}$ can be difficult. In GAMA, galaxies for which the fifth nearest neighbour is more distant than the nearest survey boundary may have erroneous environment density measurements, and the true value of $\Sigma_{5}$ is probably higher than the value measured. Galaxies for which this is a problem are more likely to be situated in the lowest density environments. We have rejected $188$ galaxies for which $\Sigma_{5}$ was not able to be reliably measured.

The GAMA catalogue also includes two other environment density estimates: A Counts In Cylinder measurement, which counts the number of galaxies in a cylinder of radius $1$ $h^{-1}$ Mpc and depth of $1000$ km s$^{-1}$, and an Adaptive Gaussian Ellipsoid (AGE) density measurement following \cite{Schawinski07}. 

Given that the results of \cite{Muldrew12} indicate that the adaptively-defined $\Sigma_{5}$ measurement will recover the underlying density field in small-scale dense environments better than the aperture-based density measurements we will use this as the metric for environment density for the majority of our analysis. We shall use the term ``local density'' to refer to $\Sigma_{5}$ and unless otherwise stated all environment density measurements will be $5^{th}$ nearest neighbour densities.

\subsection{Stellar masses}
We make use of the GAMA photometric estimates of the galaxy stellar masses derived by \cite{Taylor11}. Stellar masses were calculated using stellar population synthesis modelling of the GAMA \emph{ugriz} spectral energy distributions and assuming a \cite{Chabrier2003} stellar initial mass function. These calculations are used to produce a four-parameter fit which includes the e-folding timescale for the star formation history, age, stellar metallicity, and dust extinction. Modelling the galaxy SEDs in this fashion produces estimates of the mass-to-light ratio with typical statistical uncertainties of approximately $0.1$ dex for galaxies brighter than $r_{petro}=19.6$ mag. These measurements were made using the total integrated light for each galaxy and as such estimate the integrated stellar mass for the entire galaxy, which is often larger than the SAMI aperture.

\subsection{Sample selection}\label{Sample_selection}
While the capabilities of integral field spectroscopy allow us to construct a more complete picture of the star formation morphology of a galaxy than single fibre or long-slit spectroscopy, the nature of a large-scale hexabundle IFS survey presents us with some limitations. In particular, the SAMI instrument consists of hexabundles \citep{BlandHawthorn2011,Bryant2014} which subtend $15\arcsec$ on the sky. The combined effect of the on-site seeing and the data cube construction process results in a point spread function (PSF) full width at half maximum (FWHM) distribution with median $\sim 2\farcs 2$. Thus, we are faced with biases at both high and low redshifts. At low redshifts we encounter the problem that galaxies with higher stellar mass are not sampled out to large radii and we risk the interpretation of small-scale substructure within a galaxy as a true star formation rate gradient. Conversely, at higher redshifts galaxies will tend to have smaller angular sizes and the spatial structure can be dominated by beam smearing of the image introduced by the seeing during observation. We reduce these issues by selecting only galaxies for which the SDSS $r$-band effective radius ($r_{e}$) satisfies $0.4 r_{e} \leq 7\farcs5 \leq 3.0 r_{e}$. That is, the SAMI hexabundle field-of-view, with projected radius of $7\farcs 5$, must not sample a galaxy beyond $3.0 r_{e}$ or encompass a region of the galaxy less than $0.4r_{e}$ in radius. A total of $55$ galaxies from the original $808$ are rejected under these criteria.

\begin{figure}
\includegraphics{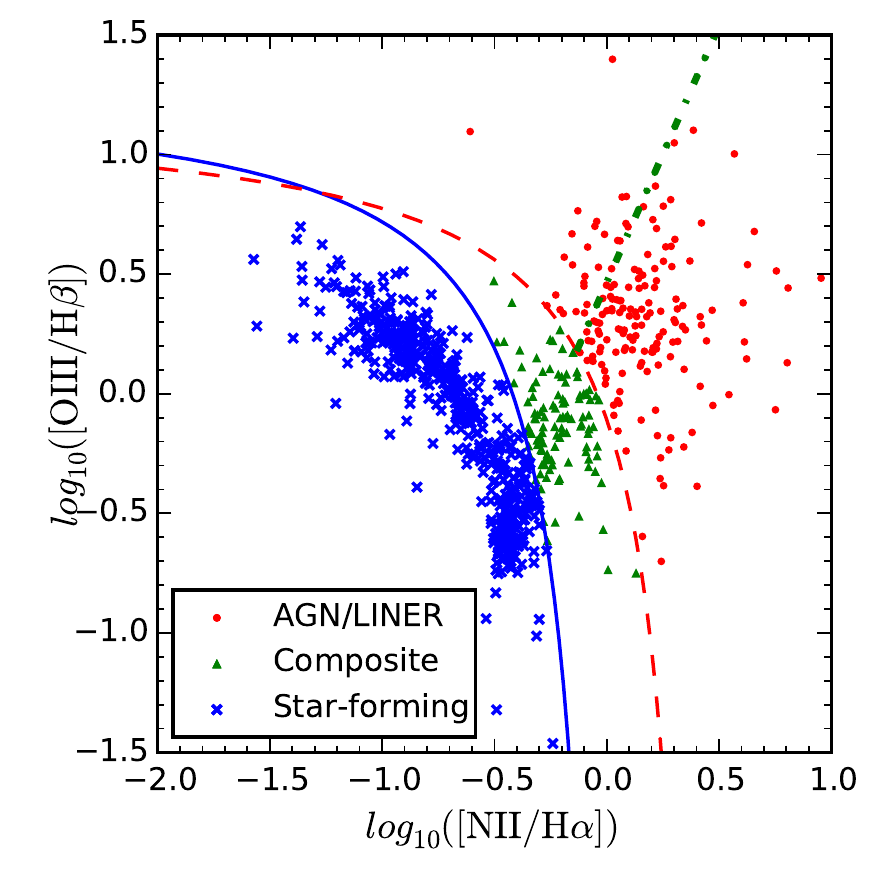}
\caption{An ionisation diagnostic diagram \protect\citep{BPT81} for the $808$ galaxies in our input sample. The red dashed curve is the AGN cutoff of \protect\cite{Kewley01} and the solid blue curve is the star-forming limit as defined by \protect\cite{Kauffmann03c}. The diagonal, green dot-dashed line separates AGN from LINERs. The emission line fluxes are extracted from the central circular $2\arcsec$ of each galaxy. Galaxies classified as AGN/LINER ($179$) are represented by red points, those with composite spectra ($111$) are marked with green triangles, and galaxies with emission line ratios consistent with ionisation from a young stellar population are classified as star-forming ($518$) are marked with blue crosses.}\label{BPT_diagram}
\end{figure}

In addition to the constraints placed on this sample by the target selection of the SAMI Galaxy Survey, some further restrictions on the galaxies analysed were required to ensure the integrity of this work. Photoionisation of ambient gas within each galaxy from non-stellar sources or an old stellar population will contaminate the measurement of star formation. This will be the case for galaxies with an active galactic nucleus (AGN) or a Low Ionisation Nuclear Emission Region (LINER). To guarantee that the measured Balmer line flux was the result of gas excitation from a young stellar population, the spectrum from a central circular $2\arcsec$ aperture was extracted from each galaxy. Within this aperture the intensities of H$\alpha$, [\ion{N}{ii}] $\lambda6583$, [\ion{O}{iii}] $\lambda 5007$ and H$\beta$ were measured and compared (see Section~\ref{linefitting} for a discussion of the emission line measurements). We used the ionisation diagnostics of \cite{Kewley01} and \cite{Kauffmann03c} to classify each central-spectrum as either AGN/LINER, composite or star-forming. Systems with line ratios above both the Kauffmann and Kewley lines were identified as AGN/LINER and those between the two diagnostic curves were classified as composite. A total of $179$ galaxies with AGN-like emission line ratios were found with this method, though the emission line signal-to-noise ratio in 76 of these were so low that this classification is uncertain. All of these galaxies were removed from the sample. A further $111$ galaxies are classified as composite. Systems classified as either AGN or composite have not been included in the final sample. A \cite{BPT81} diagnostic diagram is shown in Figure \ref{BPT_diagram} for our data and shows the separation of the star-forming sample from galaxies that are not star-forming.

In the case of edge-on disc galaxies, the radial binning technique applied to our maps cannot give an accurate picture of the true radial profile of the star formation rate. $126$ galaxies with ellipticities greater than $0.7$ were classified as edge-on disc galaxies and were therefore also rejected. We discarded an additional $199$ galaxies for which the PSF FWHM for the observation extends more than $0.75 r_{e}$ or is greater than $4\arcsec$ to minimise the effect of beam smearing on our conclusions. 
Finally, given the range of emission line strengths in the sample, it was useful to split the galaxies into two groups based on their absorption-corrected H$\alpha$ equivalent widths ($\mathrm{EW}_{\mathrm{H}\alpha}$), with positive equivalent widths indicating emission. Equivalent widths were determined by summing the entire data cube over its spatial dimensions and deriving the EW from the resulting spectrum. For this purpose, the EW is defined as the continuum-corrected H$\alpha$ flux divided by the continuum level at the H$\alpha$ wavelength extrapolated over the absorption line. As the $\mathrm{EW}_{\mathrm{H}\alpha}$ in star-forming systems is tightly correlated with their specific star formation rate, we define galaxies for which the integrated aperture spectrum has $\mathrm{EW}_{\mathrm{H}\alpha} >1$ \AA ~in emission as star-forming and galaxies with $\mathrm{EW}_{\mathrm{H}\alpha}\leq 1$ \AA ~as quiescent. This distinction separates those galaxies in the star-forming ``main sequence" from those which appear to be quenched. \cite{CidFernandes11} recommend a $3$ \AA ~cut in $\mathrm{EW}_{\mathrm{H}\alpha}$ to separate passive and star-forming galaxies. However, this recommendation was based on the use of SDSS single-fibre spectroscopy and is thus sensitive to the distribution of star formation within the system. Our summation over the entire SAMI aperture ensures that galaxies with, for example, passive centres and star-forming edges are not selected against. The reduction of the $\mathrm{EW}_{\mathrm{H}\alpha}$ cut takes into account the contribution from the passive regions of the galaxy. 

We find that the rejection of galaxies based on their central spectrum, $\mathrm{EW}_{\mathrm{H}\alpha}$ and ellipticity is sufficient to eliminate all significant non-central non-star-forming emission from our sample. Highly inclined galaxies that show extra-planar emission or shock-excited winds driven by star formation \citep[see e.g.][]{Ho16} are rejected. For those that are not inclined to our line of sight, the contribution to the total flux is generally low enough as to not be detected. Galaxies with spatially extended line emission excited by old stellar populations \citep{Sarzi2010,Belfiore2016} do not satisfy the $\mathrm{EW}_{\mathrm{H}\alpha}$ criterion.

\begin{figure}
\includegraphics{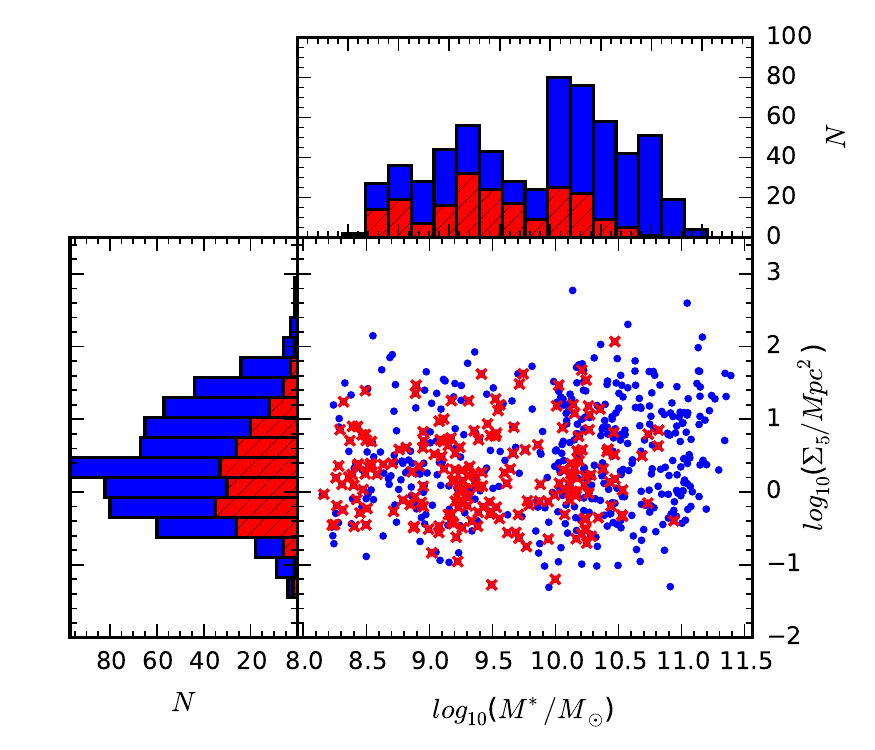}
\caption{The distribution of galaxy stellar masses and fifth-nearest-neighbour surface densities for our sample. Blue and red points represent the entire sample of $808$ observed SAMI galaxies and red indicates the sample after the application of the constraints outlined in Sections~\ref{target_selection} and \ref{Env_density}. The upper histograms show the distributions of $\log_{10}(\rm{M}_{*}/\rm{M}_{\odot})$ and the histograms on the left are the distributions of $\log_{10}(\Sigma_{5}/\rm{Mpc^{2}})$ with blue for the input sample and red hatched for the final star-forming sample. Note that the subsample retained for analysis has reduced coverage of the $\rm{M}_{*}$-$\Sigma_{5}$ parameter space, with most galaxies lost at high stellar mass and in high density environments.}\label{m_env_dist}
\end{figure}

\begin{figure*}
\includegraphics{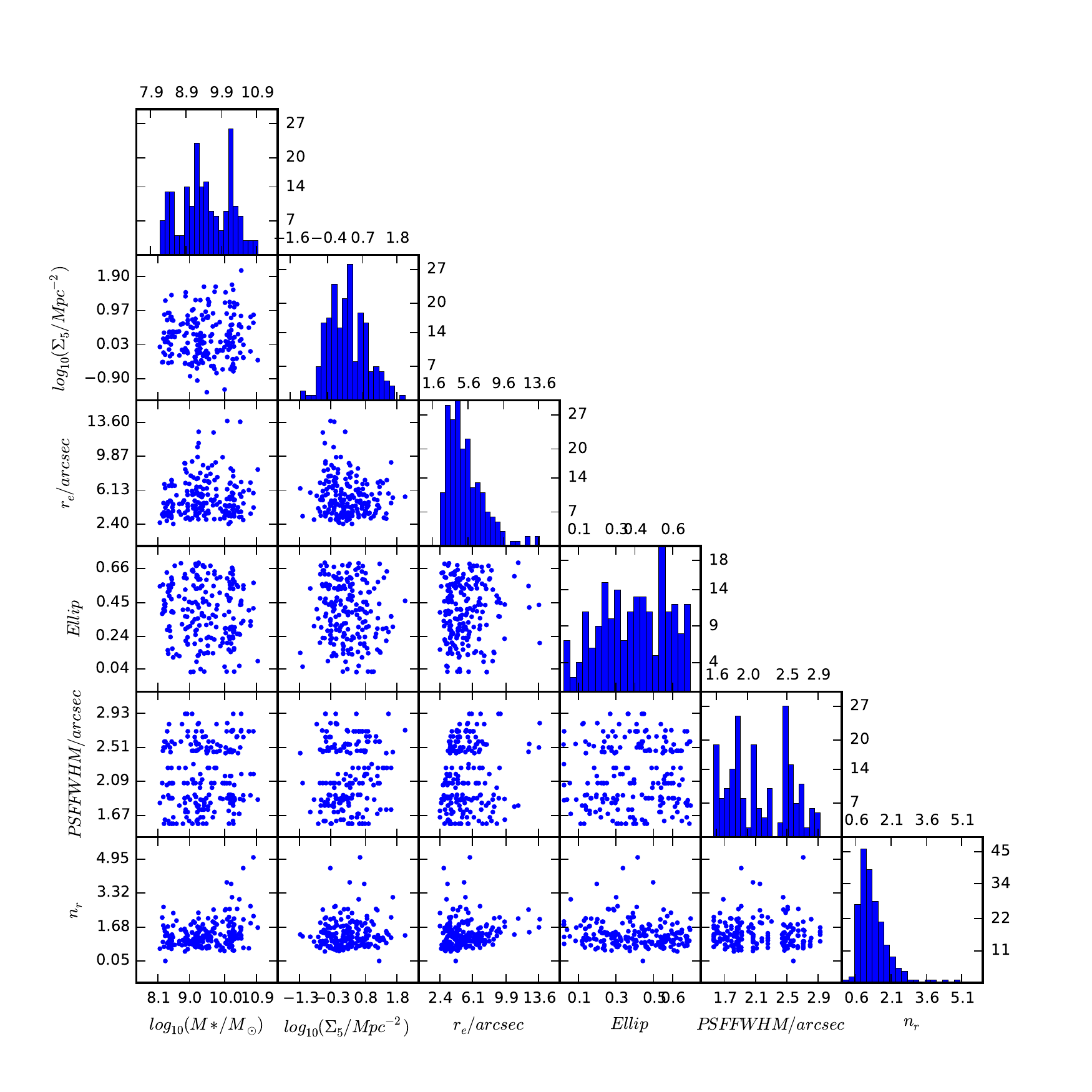}
\caption{Global galaxy parameters for the final sample of \SFNUM\ galaxies. We show the distributions of $\log_{10}(\rm{M}_{*}/\rm{M}_{\odot})$, $\log_{10}(\Sigma_{5}/Mpc^{-2})$, $r_{e}$, ellipticity, PSF FWHM and the S{\'e}rsic $n_r$. Along the diagonal we show the histograms of each of these parameters while the off-diagonal diagrams show the relationship between the variables for each galaxy.}\label{AllParams}
\end{figure*}

Our final star-forming sample comprises \SFNUM\ galaxies. Note that a number of galaxies that were rejected failed on multiple criteria. The star-forming sample contains galaxies with stellar masses in the range $10^{8.2}$-$10^{10.9} \:  \mathrm{M}_{\odot}$ and fifth nearest neighbour environment densities in the range $10^{-1.3}$-$10^{2.1} \: \mathrm{Mpc}^{-2}$. This range of environment densities incorporates galaxies from low-density field environments to groups of halo mass $10^{14.5}$ M$_{\odot}$ and between $2$ and $104$ members per group, though the current sample does not include all the galaxies within each group. The variation of the star-forming properties of galaxies in groups will be the subject of a future paper. The distributions of galaxy stellar masses and local environment densities are displayed in Figure~\ref{m_env_dist}. In this figure, blue colours indicate the input sample of $808$ galaxies and red indicates galaxies that remain after the above constraints are applied. There is a dearth of star-forming galaxies with stellar masses above $\mathrm{M}_{*}=10^{11}\, \mathrm{M}_{\odot}$. This is consistent with mass and environment quenching as described by \cite{Peng2010}. We also show the distributions of other relevant parameters in our final sample in Figure \ref{AllParams}. This final sample covers a wide range of galaxy stellar masses, environment densities and morphologies.

\section{Spectral fitting and analysis} \label{DataReduction}

\subsection{Binning And The Balmer Decrement}\label{binning}
Obtaining accurate measurements of emission line fluxes, especially H$\beta$, is essential to estimating and properly correcting for the presence of dust obscuration along the line of sight. This is particularly important when the lines are observed to be only weakly in emission. In order to properly account for the underlying stellar absorption features, a S/N of at least $10$ per angstrom in the continuum is desirable. In regions of galaxies with low surface brightness, such as the outer edges of the disk, some level of spatial binning is often required to achieve this. An added complication is that dust attenuation along the line of sight is corrected for using a non-linear combination of H$\alpha$ and H$\beta$ fluxes. In each spatial bin, the observed H$\alpha$ fluxes, $f_{\mathrm{H}\alpha}$, are corrected for dust attenuation along the line of sight according to the \cite{Cardelli89} dust extinction law. This correction uses the deviation of the Balmer decrement ($BD$; the ratio $f_{\mathrm{H}\alpha}/f_{\mathrm{H}\beta}$) from the theoretical value of 2.86 for case B recombination to model the amount of intervening dust along the line of sight. 
\begin{equation}
F_{\mathrm{H}\alpha}=f_{\mathrm{H\alpha}}\left(\frac{BD}{2.86}\right)^{2.36}
\end{equation}

\begin{figure}
\includegraphics[width=\columnwidth]{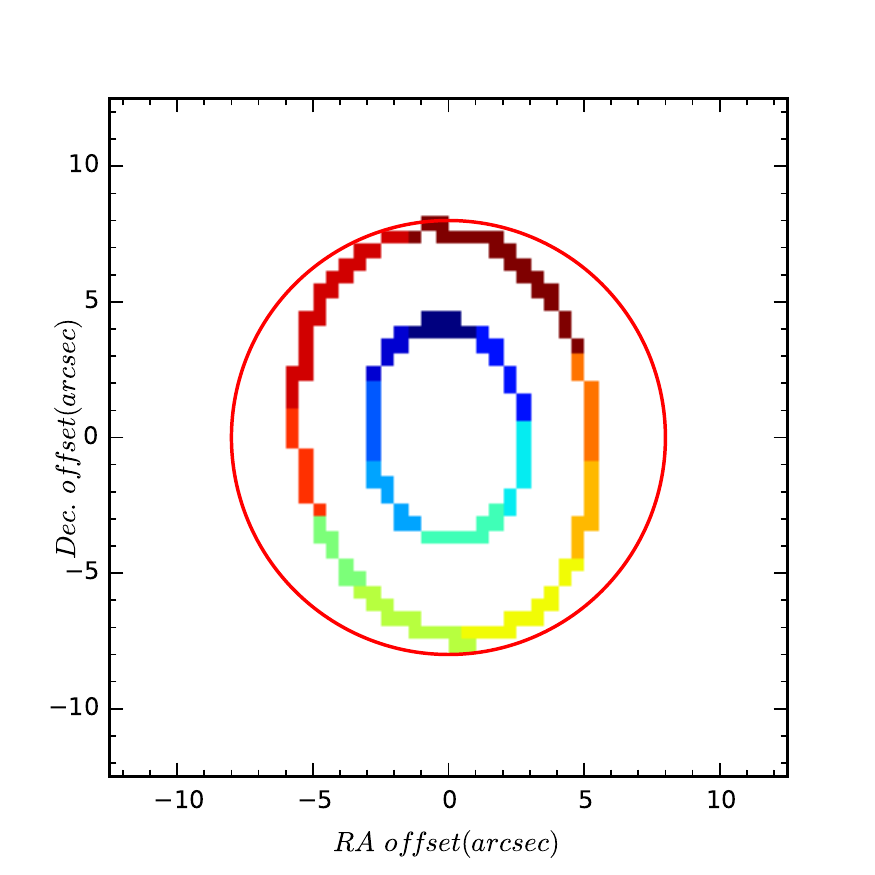}
\caption{An example of Annular-Voronoi binning which has been applied to the SAMI data cubes. In this example, spaxels in two annuli with an ellipticity of 0.28 at a position angle of $98.8\degree$ and a width of one spaxel are grouped into contiguous sub-bins, represented here as groups of pixels with the same colour. Spectral fitting, emission line integration and dust corrections are performed independently within each sub-bin. The red circle represents the size of the SAMI aperture. For clarity, only two annular bins are shown, although the entire SAMI field is partitioned into bins.}\label{bins}
\end{figure}

This extinction correction assumes the dust to be a foreground screen that is not cospatial with the emission nebulae \citep{Calzetti01}. The addition of fluxes from spatially-distinct regions of a galaxy will result in an incorrect dust correction and an underestimation of the corrected H$\alpha$ flux. With these constraints in mind we have implemented a modified version of the adaptive binning algorithm of \cite{CappellariAndCopin03}, which is based on the Voronoi tessellation of bins within an image to achieve a desired S/N. For this work we have performed what we term `Annular-Voronoi binning'. In this scheme we define Voronoi bins within a series of concentric elliptical annuli with ellipticities and position angles defined by the r-band morphology of the target galaxy. Within each annulus a number of sub-bins are constructed to achieve a target S/N of $10$ per \AA ~in the continuum in a $200$ \AA-wide window around the redshifted wavelength of the H$\beta$ line. The construction of each bin is subject to the constraint that the contributing spaxels must be contiguous. The determination of the variance value in each Voronoi bin takes into account the spatial covariance in the SAMI data described by \cite{Sharp15}. The binning is applied to both the blue and the red SAMI data cubes. This scheme has the advantage of improving the reliability of continuum subtraction, retaining the radial structure within galaxies and maintaining the locality of the dust corrections. An example of such an annular bin is shown in Figure  \ref{bins}. 

We have calculated the H$\alpha$ luminosity, L(H$\alpha$), of each galaxy using both a single global dust correction from the spatially integrated IFU ($SFR_{glob}$) and from summing locally dust corrected H$\alpha$ within the annular Voronoi bins ($SFR_{loc}$). The star formation rate can be calculated from the total L(H$\alpha$) using the \cite{Kennicutt98} relation with a \cite{Chabrier2003} IMF:
\begin{equation}
\mathrm{SFR}=\frac{L(H\alpha)~(W)}{2.16\times 10^{34}} ~\mathrm{M}_{\odot} ~\mathrm{yr}^{-1}
\end{equation}

\begin{figure}
\includegraphics{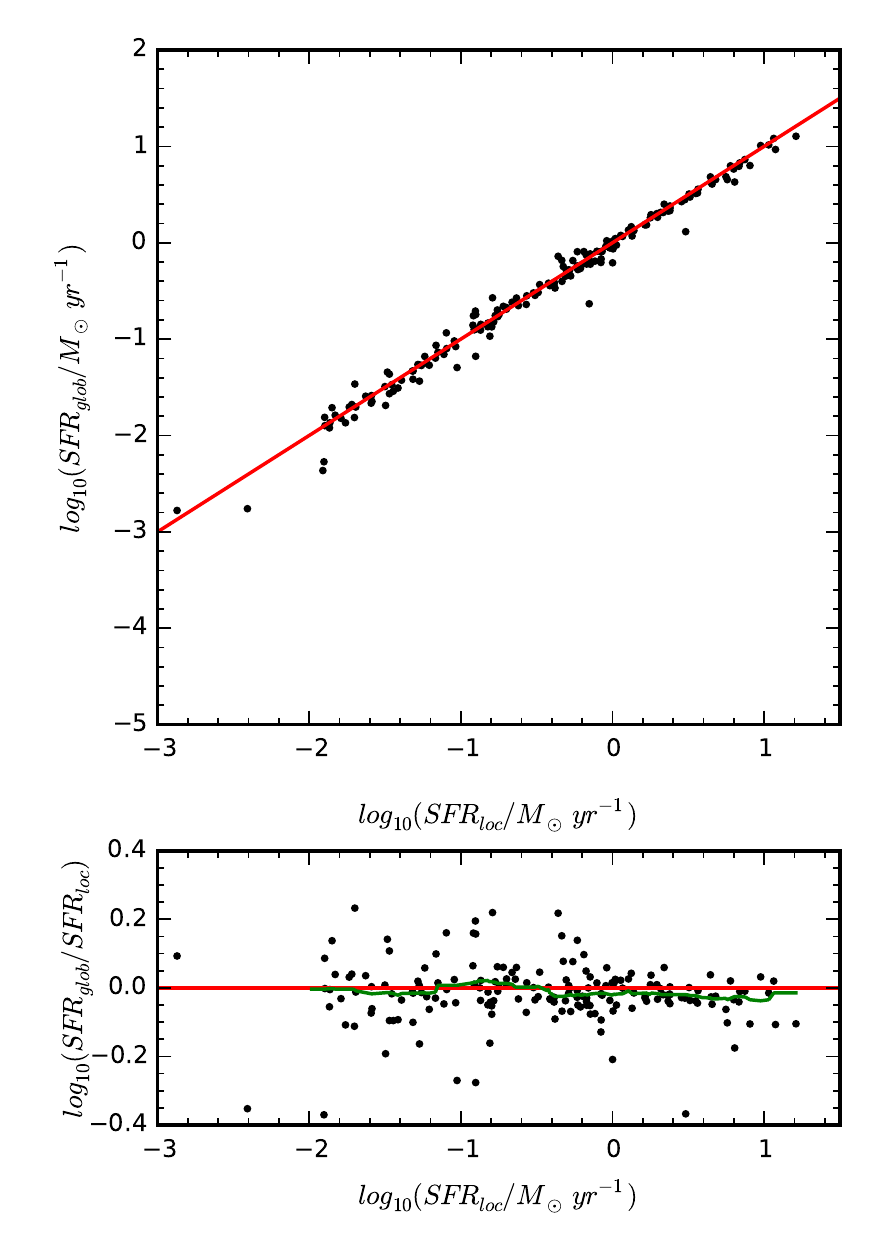}
\caption{Comparison of integrated star formation rates for objects in the star-forming sample using local dust corrections within each Annular-Voronoi bin and with a single dust correction for the whole galaxy. Red lines indicate the line of equality. In the upper panel we compare these two methods. In the lower panel we show the log ratio of the star formation rate estimate for these two methods. 
The green line is the running median, calculated in a sliding window of width $0.5$ dex. There is a shallow downward trend, indicating the global dust correction is underestimating the star formation rate for the most star-forming galaxies. For galaxies forming stars at a rate greater than $1$ M$_{\odot}$ yr$^{-1}$, the average $SFR_{glob}/SFR_{loc}=0.917 \pm 0.016$.}\label{SFRcompare}
\end{figure}

It should be noted that we have not corrected the star formation rate for aperture effects. Despite the ability of integral field spectroscopy to sample a large area of a target source, we are often unable to observe H$\alpha$ emission in the outer regions of galaxies. A correction for this aperture bias has been developed by \cite{Hopkins03} and \cite{Brinchmann04} for single fibre spectroscopic observations of galaxies but an analogous correction has not been applied to this data \citep[but see][for a discussion of aperture corrections to star formation rates with SAMI]{Richards16}. There is no systematic correlation between the projected sizes of the galaxies in our sample and other galaxy properties including $\mathrm{M}_{*}$ and $\Sigma_{5}$, meaning that aperture effects will not have a strong impact on the integrated specific star formation rates we present here. We compare the integrated star formation rates (SFRs) for the star-forming galaxies in our sample using the two different dust correction methods in the upper panel of Figure \ref{SFRcompare}. A one-to-one relationship appears to hold for the star formation rates of galaxies in the star-forming sample. For quiescent galaxies neither measurement is very accurate as the relative errors become high. Furthermore, the lower panel of Figure~\ref{SFRcompare} shows that above a SFR of $1$ M$_{\odot}$ yr$^{-1}$, the global dust correction underestimates the SFR by approximately $8 \%$ on average. This discrepancy is as expected and we thus use the locally dust corrected star formation rates for the remainder of this work. 

\subsection{Emission line fluxes}\label{linefitting}
\begin{figure*}
\includegraphics[trim=0.0cm 0.5cm 0cm 1cm,clip]{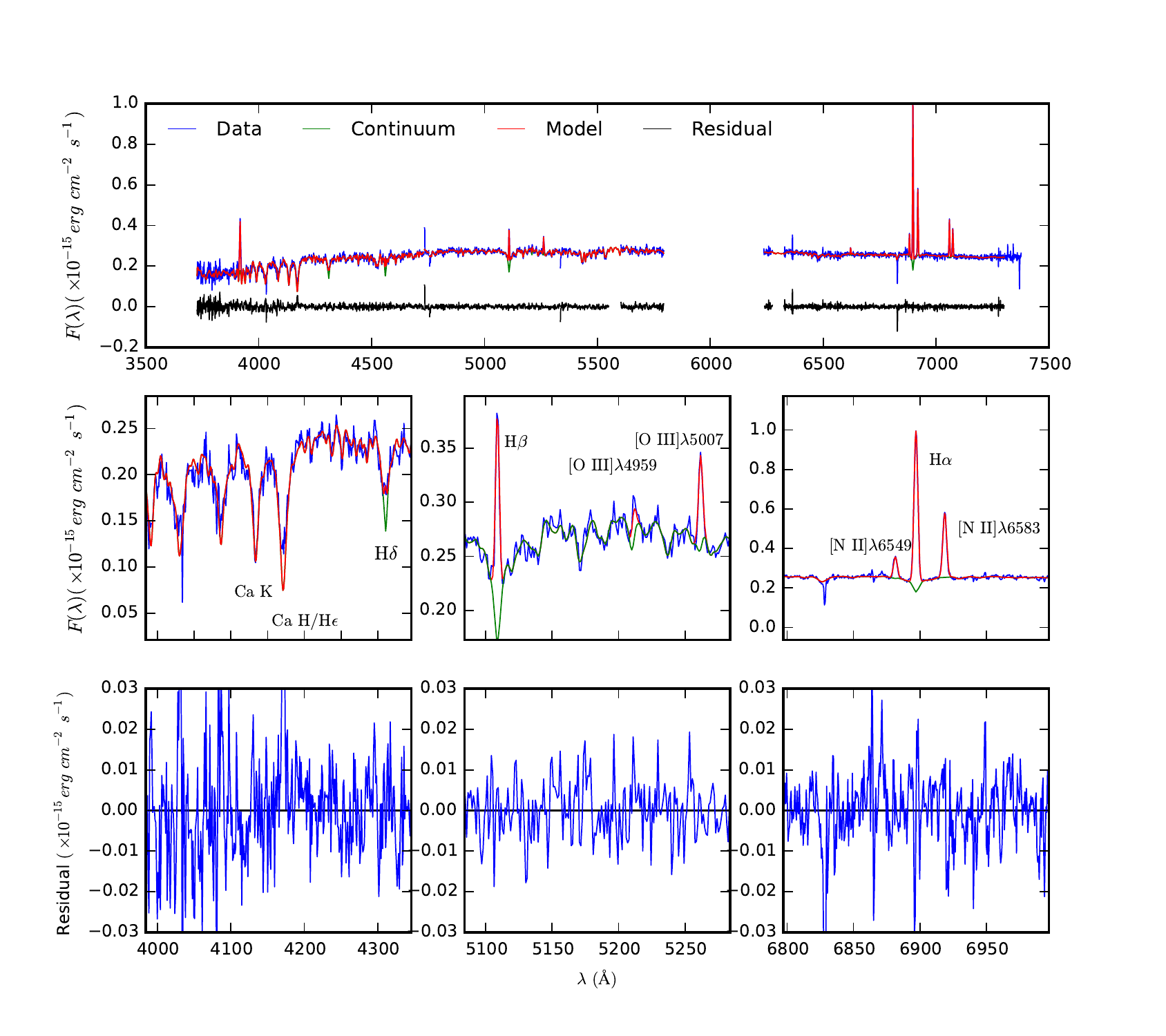}
\caption{An example of spectral fitting using \textsc{lzifu} for a single spaxel in a star-forming galaxy (GAMA catalogue ID: $517070$). This is a typical star-forming galaxy and was chosen for illustrative purposes based on the intermediate S/N of the continuum and emission lines so that all the spectral features are easily visible within a single panel. The fit includes a linear combination of stellar templates to model the continuum and a series of Gaussians for the emission lines. In the upper row we show the spectrum, fit and residuals to the entire optical spectrum for the galaxy in a single spaxel. The central row shows the spectrum continuum and emission line fit around the D$_{n}4000$, H$\beta$ and H$\alpha$ spectral features. The lower row shows the residuals to the fit in the corresponding wavelength range. The horizontal black lines denote the zero point. Higher order Balmer emission lines have not been fitted, as only H$\alpha$, H$\beta$ and H$\delta$ were used for this work. The spikes in the residual spectrum in the first column are the result of these emission lines not being accounted for in the model.}\label{spec_fit}
\end{figure*}

Fluxes were extracted from each annular Voronoi bin using a modified version of \textsc{lzifu}, a pipeline designed by \cite{Ho14} and described in detail by \cite{Ho2016b}. \textsc{lzifu} incorporates stellar template fitting to the continuum using Penalized Pixel Fitting \citep[pPXF;][]{Cappellari04} and returns a multiple component Gaussian fit to the emission lines using the \textsc{mpfit} algorithm \citep{Markwardt09}. The accuracy of the emission line flux measurements is ultimately tied to the accuracy with which we can fit and subtract the underlying stellar absorption features. The stellar continuum in each spectrum has been modelled using pPXF to fit a combination of simple stellar population models (SSP) from the MILES library \citep{Vazdekis10} as well as an 8$^{th}$ degree multiplicative polynomial to account for any residual flux calibration errors and the reddening effect of dust. Strong emission lines were masked during the fitting process and weaker emission lines were clipped by setting the \textsc{clean} keyword when invoking pPXF. A $65$ template subset of the full MILES SSP library was used to independently fit the stellar continua in each spatial bin. These templates span $5$ stellar metallicities in the range $-1.71 \leq  \mathrm{[Z/H]} \leq 0.22$ and for each of these metallicities sample SSPs at $13$ logarithmically spaced ages between $0.063$ and $14$ Gyrs. These models were formed at a spectral resolution of $2.5$ \AA ~FWHM. Note that this is higher than the spectral resolution of the 580V grating used in the blue arm of the SAMI instrument but lower than the spectral resolution of the $1000$R grating used for the red arm of SAMI. This discrepancy in the resolutions will mean that a minimum stellar velocity dispersion can be measured from our data. At $6890.94$ \AA, the wavelength of the H$\alpha$ absorption line at a redshift of $0.05$, this corresponds to a minimum measurable stellar velocity dispersion of $37.8$ km s$^{-1}$. This velocity dispersion is a reasonable lower limit on the measured dispersion of spectra in SAMI. The values for the velocity dispersion returned by the pPXF fit of the MILES templates to the red SAMI data cubes are unreliable, however, a good fit to the data is still obtained and the stellar absorption correction is valid. In any case, the emission line fitting is done after the continuum has been subtracted from the spectrum and the spectral resolution of the stellar templates has no effect on the emission line measurements. In each spaxel the wavelengths of the emission lines fit are allowed to float relative to the redshift of the stellar continuum, but are fixed with respect to each other. \textsc{lzifu} outputs maps for the distribution of the strongest emission lines which have been corrected for absorption by the stellar continuum. An example of the resulting fit to a spectrum that shows all the relevant features is shown in Figure \ref{spec_fit}.

\subsection{Radial Profiles}\label{Method}
\begin{figure*}

\includegraphics[trim=0.5cm 0.5cm 0.5cm 0.5cm, clip,scale=0.9]{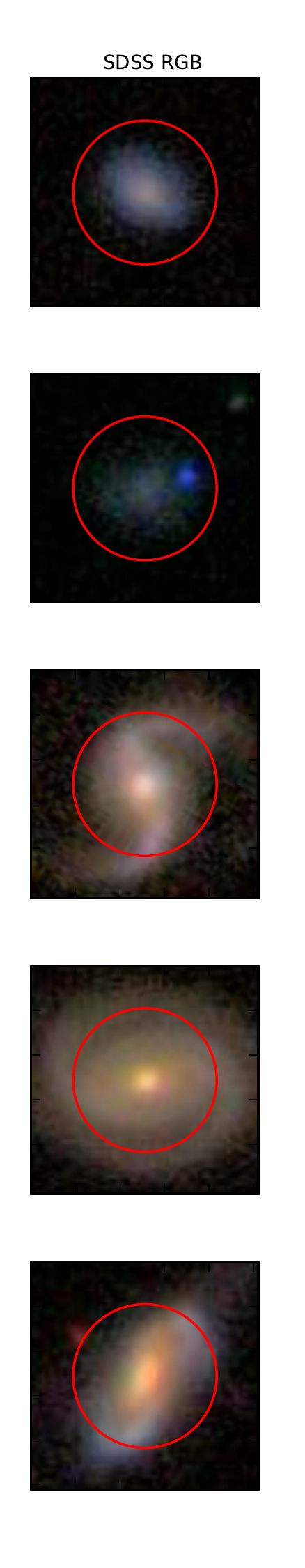} \includegraphics[trim=0.0cm 0.5cm 0.0cm 0.5cm, clip,scale=0.9]{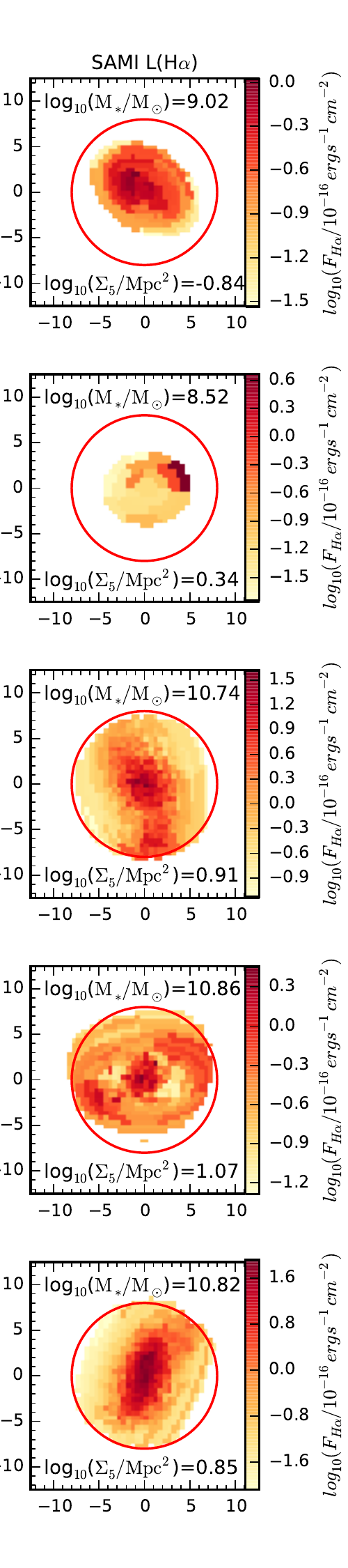}  \includegraphics[trim=0.0cm 0.5cm 0.0cm 0.5cm, clip,scale=0.9]{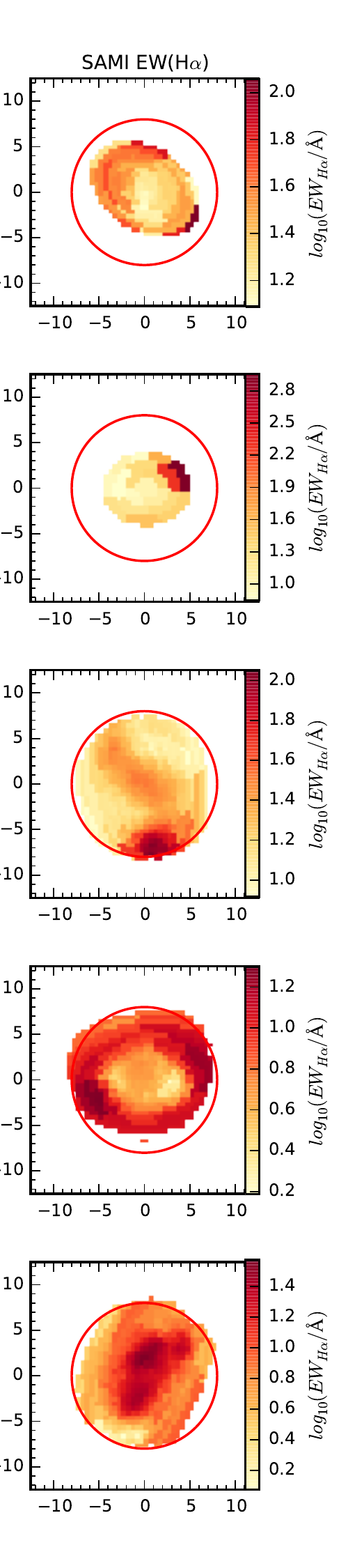}  \includegraphics[trim=0.0cm 0.5cm 0.0cm 0.5cm, clip,scale=0.9]{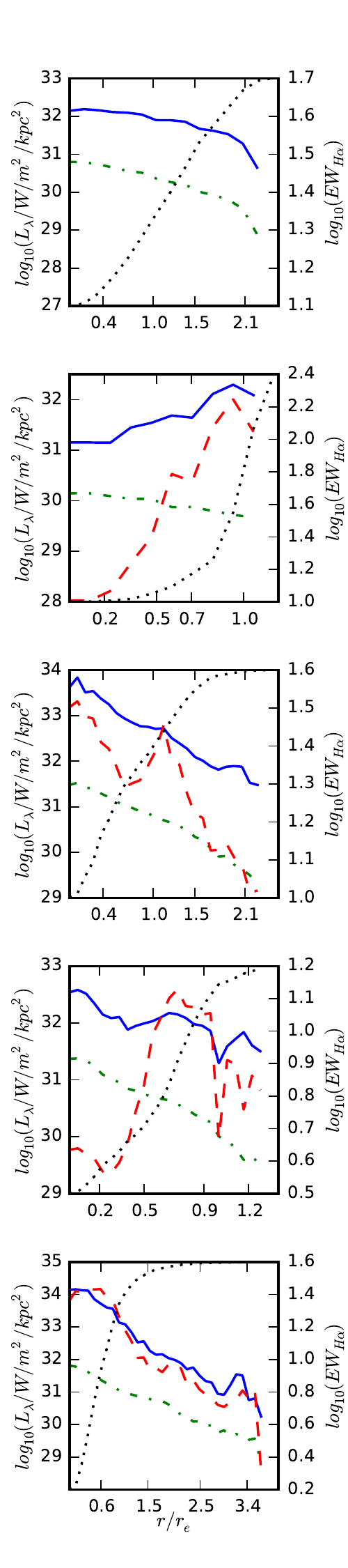}

\caption{Example galaxies with a range of stellar masses, local environment densities and H$\alpha$ morphologies. On the far left is a colour optical image from the SDSS $gri$ photometry, left-of-centre is the extinction-corrected $\mathrm{\log}_{10}$(H$\alpha$) flux map reconstructed from the annular-Voronoi bins and right-of-centre is the H$\alpha$ equivalent width map, made from the annular-Voronoi bins. The SDSS image and each SAMI map are located in boxes that are $25\arcsec$ on each side. In each image the red circle marks the $15\arcsec$ SAMI field of view. On the far right we show the radial profiles for $\log_{10}(\mathrm{H}\alpha / W /kpc^{2})$ luminosity surface density (blue solid), $\log_{10}(\mathrm{EW}_{\mathrm{H}\alpha})$ (red dashed), the SAMI red arm integrated continuum (green dot-dashed) and the curve of growth for dust-corrected H$\alpha$ (black dotted). Each curve of growth is scaled such that $0\%$ of the flux from the galaxy is at the bottom of the panel and $100\%$ of the flux is at the top.}
\label{rpeg} 
\end{figure*}

The radial profiles of star formation were calculated from the \textsc{lzifu} outputs. Each galaxy was treated as a circular disc tilted at some angle to the line of sight. To take into account the effect of this projection onto the sky, the radial profiles of H$\alpha$ were constructed by averaging the dust corrected fluxes over a series of concentric elliptical annuli. These annuli were centred on the centroid of the continuum flux of each galaxy, with their ellipticities and position angles based on the GAMA photometric model fits to reanalysed SDSS DR7 $r$-band images \citep{Kelvin12}. Some examples of radial profiles, dust-corrected H$\alpha$ maps and EW(H$\alpha$) maps are shown for galaxies with a range of stellar masses and star-forming morphologies in Figure \ref{rpeg}. Taking the inclination of the galaxy to our line of sight improves our estimate of the radial distribution of H$\alpha$. However, in some cases the presence of strong morphological irregularities such as bars, and spiral arms can skew the photometric fit and yield erroneous parameters. A visual inspection of the GAMA S{\'e}rsic profile fits for the \SFNUM star-forming galaxies in the sample resulted in the identification of $15$ galaxies where the galaxy morphology skews the photometric fit and the measured ellipticity does not constrain the inclination to our line of sight well. The prevalence of these galaxies shows no trend with stellar mass or environment density. Moreover, this effect changes the measured average surface brightness at $1 \, r_{e}$ typically by less than $0.2$ dex and never more than $0.4$ dex. We therefore conclude that such galaxies do not affect the results of this work.

\begin{figure*}
\includegraphics{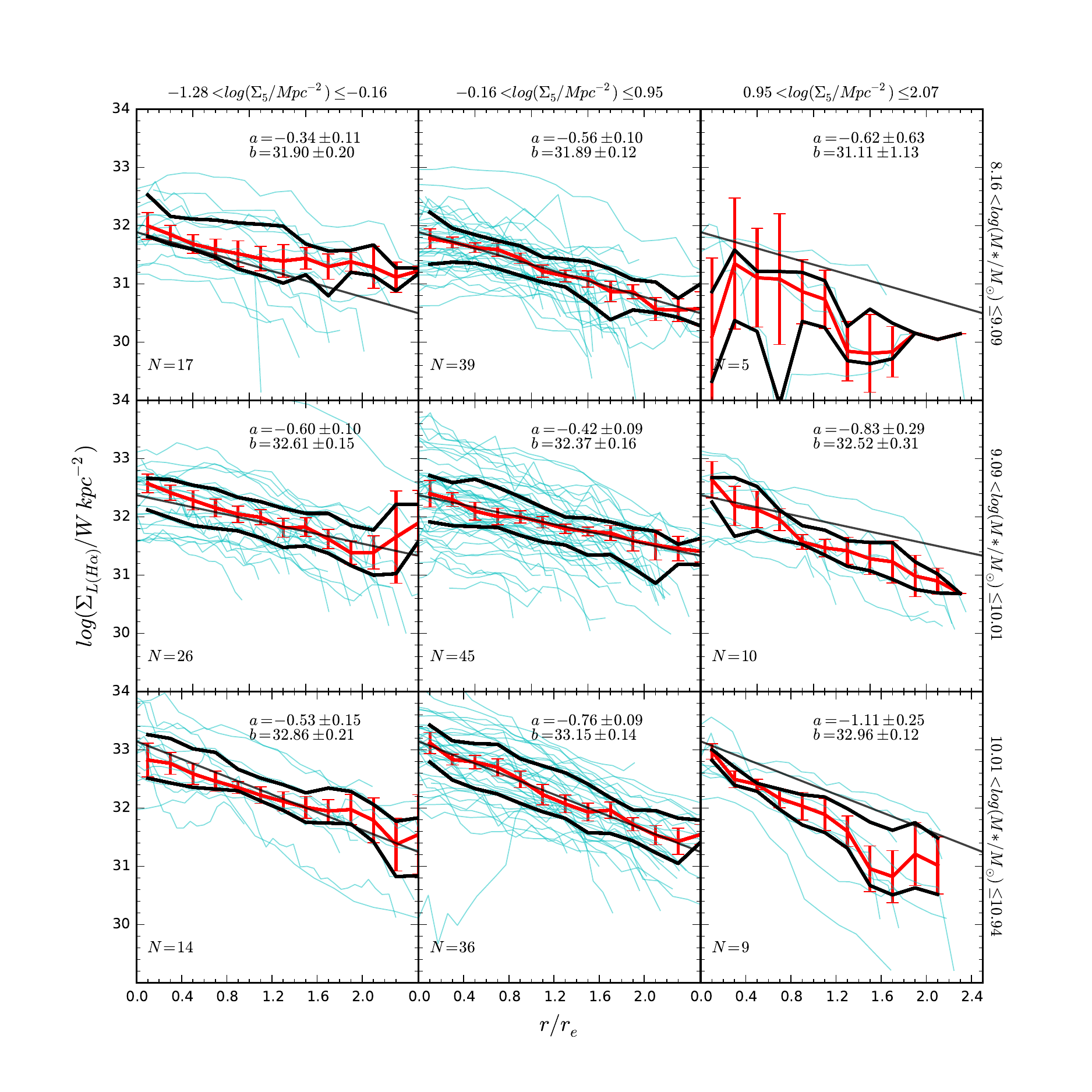}
\caption{The radial profiles of H$\alpha$ luminosity surface density for each of the star-forming galaxies in the sample, split into three bins of stellar mass (increasing from top to bottom) and environment density (increasing from left to right). Each bin covers one third of the total range in $\log_{10}(\mathrm{M}_{*}/\mathrm{M}_{\odot})$ and $\log_{10}(\Sigma_{5}/\mathrm{Mpc}^{2})$. The profile of each galaxy is shown in cyan, with the median profile of each bin shown in red. Thick black lines show the $25^{th}$ and $75^{th}$ percentiles of the radial profiles. For each stellar mass interval an exponential fit to the median profile of the central density bin is overplotted in black on each panel for comparison. In the upper right of each panel we show the best-fit parameters of the exponential fit (using Equation \ref{Exponential}) to the median profile in that bin of stellar mass and environment density.}\label{Profile_Grid}
\end{figure*}

For an ensemble of galaxies it will be useful to construct an average radial profile that is indicative of the radial behaviour of the star formation distribution for the group as a whole. We can construct a `median profile' for a selection of galaxies by choosing the median value of log-flux in some radial bin. For this work we have rebinned each radial profile onto a grid with the radial axis in units of $r_{e}$, the $r$-band effective radius within which one half of the total $r$-band light is contained. Each radial bin has a size of $0.2 \, r_{e}$, and in each bin we calculate the value of the median radial profile by simply extracting the median flux. At each point the errors on the measurement of the median are computed from the standard deviation of $1000$ median values obtained by bootstrap resampling of radial profiles in the sample. The radial profiles of H$\alpha$ for all the star-forming galaxies in our sample are shown in Figure \ref{Profile_Grid}. These profiles are split into three equally sized bins of $\log_{10}(\Sigma_{5}/\mathrm{Mpc}^{2})$ and $\log_{10}(\mathrm{M}_{*}/\mathrm{M}_{\odot})$. To quantify the changes in the star formation distribution in galaxies of different stellar masses and in different environments we fit each median profile in Figure \ref{Profile_Grid} with an exponential of the form
\begin{equation} \label{Exponential}
\mathrm{log_{10}}(\Sigma_{L(\mathrm{H}\alpha)})= a \times r/r_{e} + b.
\end{equation}

The parameters of these fits are displayed in Figure \ref{Profile_Grid} at the upper right of each panel. The errors on these parameters are obtained by bootstrap resampling the galaxies in each stellar mass and environment density bin, recalculating and refitting the median profiles.

\subsection{D$_{n}4000$ and H$\delta_{A}$ gradients}
The strength of the D$_{n}4000$ index in a galaxy spectrum is indicative of the age of the stellar population. While the D$_{n}4000$ index alone is insufficient to determine the precise age of the stellar population, particularly where the most recent episode of star formation was more than 1 Gyr ago \citep{Kauffmann03a}, gradients of its strength are indicative of underlying stellar population age gradients. We have measured the D$_{n}4000$ index, as defined by \cite{Balogh99}, in concentric elliptical annuli around the centres of the galaxy.
Gradients of the D$_{n}4000$ index are calculated from the values at the centre and within an ellipse with a semi-major axis length of $5 \arcsec$.
Due to the stochastic occurrence and short lifetimes of individual \ion{H}{ii} regions in galaxies, the D$_{n}4000$ gradients can be used to confirm that any radial trend in the H$\alpha$-derived star-forming properties of a galaxy can be explained as genuine quenching rather than as random, short-timescale fluctuations in the H$\alpha$ distribution. In particular, positive D$_{n}4000$ gradients (i.e. D$_{n}4000$ stronger in the galaxy outskirts) are evidence for outside-in quenching acting or the recent enhancement of centrally located star formation in a formerly passive galaxy. 

The H$\delta_{A}$ absorption line equivalent width is also useful in estimating the age of a galaxy's stellar population, since Balmer absorption lines are strongest in relatively short lived B and A-type stars. We measure the strength of the H$\delta$ absorption from the spectrum after subtraction of the emission line fit using the index bands defined by \cite{WortheyOttaviani1997}. A gradient is then calculated as for D$_{n}4000$.

\section{Results}\label{Results}
In this section we present our analysis of the H$\alpha$ emission in the star-forming galaxies observed by SAMI. This includes both the integrated star formation rates and the spatial distribution of star formation in the galaxies as a function of their stellar mass and their local environment density.

\subsection{Integrated Star Formation Rates}\label{integrated_sfr}

\begin{figure*}
\includegraphics{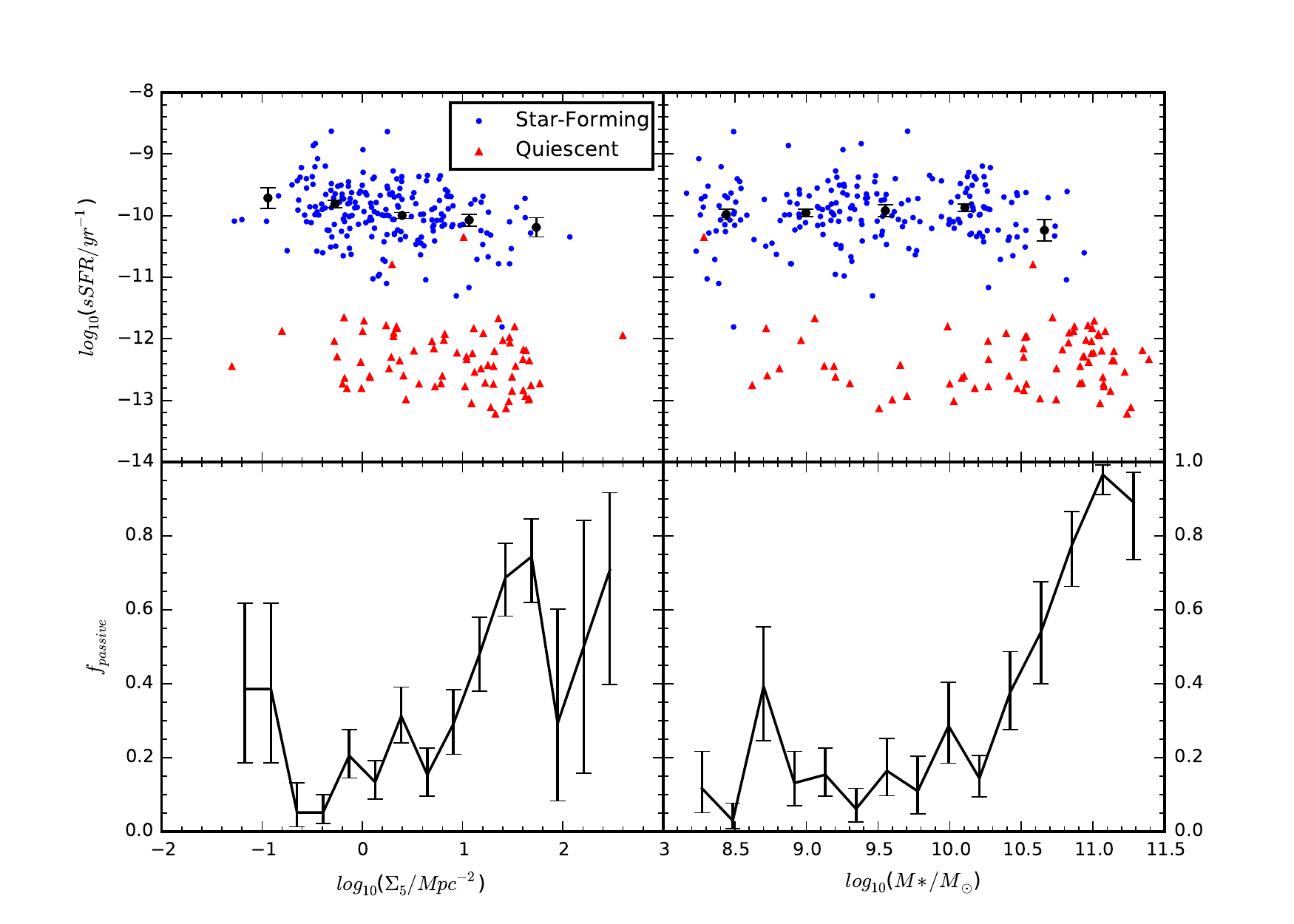}
\caption{The upper row shows the specific star formation rates of galaxies as a function of their environment density (left) and stellar mass (right). Data for star-forming galaxies are shown with blue points, while the data for the passive galaxies are shown with red triangles. The black points are the median values of the specific star formation rates of star-forming galaxies in equally spaced bins of $log_{10}(\Sigma_{5}/\mathrm{Mpc^{2}})$ or $log_{10}(M_{*}/\mathrm{M_{\odot}})$. The errors are estimated as the standard deviation of the median in $1000$ samples bootstrapped from each bin.
The lower row shows the expectation value of the fraction of galaxies that we classify as quiescent as a function of $\Sigma_{5}$ and $\rm{M}_{*}$. The passive fraction of galaxies increases above $\rm{M}_{*}=10^{10}$ M$_{\odot}$ and above $\log(\Sigma_{5}/\rm{Mpc^{2}})=0.5$.}\label{ssfr_m_env_frac}
\end{figure*}

We investigate the effect of local environment density on the integrated star formation in galaxies and see in Figure~\ref{ssfr_m_env_frac} that the overall level of star formation in all galaxies (star-forming and passive) decreases for galaxies in higher density environments. This is in agreement with other surveys \citep[e.g.][]{Balogh04,Wijesinghe12}. The relationship between the specific star formation rates ($\rm{SFR}/\rm{M}_{*}$; sSFR) of star-forming galaxies and their local environment density and stellar mass can be seen in Figure~\ref{ssfr_m_env_frac}. In the left panel of the upper row we compare sSFR to $\Sigma_{5}$. There is evidence for a correlation between the sSFR of star-forming galaxies and $\Sigma_{5}$ ( Spearman's $\rho=-0.27$, $p=1.3\times 10^{-4}$) though again, a higher fraction of quiescent systems exist in high-density regions. The passive fraction of galaxies increases both with stellar mass and environment density, with a steady rise in the passive fraction above $\log_{10}(\Sigma_{5}/\rm{Mpc}^{2}) = 0.5$, corresponding to a fifth-nearest neighbour separation of $0.7$ Mpc. In the right panel of the upper row in Figure~\ref{ssfr_m_env_frac} we compare the sSFR to the stellar mass of each galaxy. There is no correlation between the specific star formation rate and the stellar masses of star forming galaxies in our sample, with a Spearman rank correlation coefficient of $\rho=-0.016$ and $p=0.83$. The quiescent fraction in galaxies appears to increase abruptly above $\sim 10^{10}$ M$_{\odot}$, in accordance with previous studies \citep[e.g.][]{Kauffmann2003b,Geha2012}. 
The quiescent fraction of galaxies as a function of mass and environment density given the data is displayed in the lower row of Figure~\ref{ssfr_m_env_frac}. The fractions here are calculated as the $50^{th}$ percentile of the beta distribution defined by the total number of galaxies in each bin and the number of these that are quiescent, with the lower and upper error bars being the $32^{nd}$ and $68^{th}$ percentiles of this distribution respectively. This approach is taken following \cite{Cameron11}. We see that the passive fraction of galaxies increases above $\log_{10}(\Sigma_5/\rm{Mpc}^{2})=0.5$ and above masses of $\log_{10}(\rm{M_{*}}/\rm{M}_{\odot})=10$.

The correlation between specific star formation rate and local environment density is weak but significant for all star-forming systems. The dominant trend appears to be the changing fraction of passive galaxies in higher density regions, particularly at higher stellar mass, though a weak trend may be evident for the star-forming galaxies. This conclusion is strongly dependent on the definition of a passive galaxy and our ability to measure star formation rates in systems with weak emission lines. This result is largely consistent with the results of \cite{Wijesinghe12} who found that the changing fraction of passive galaxies dominates the environmental trend.

\begin{figure}
\includegraphics{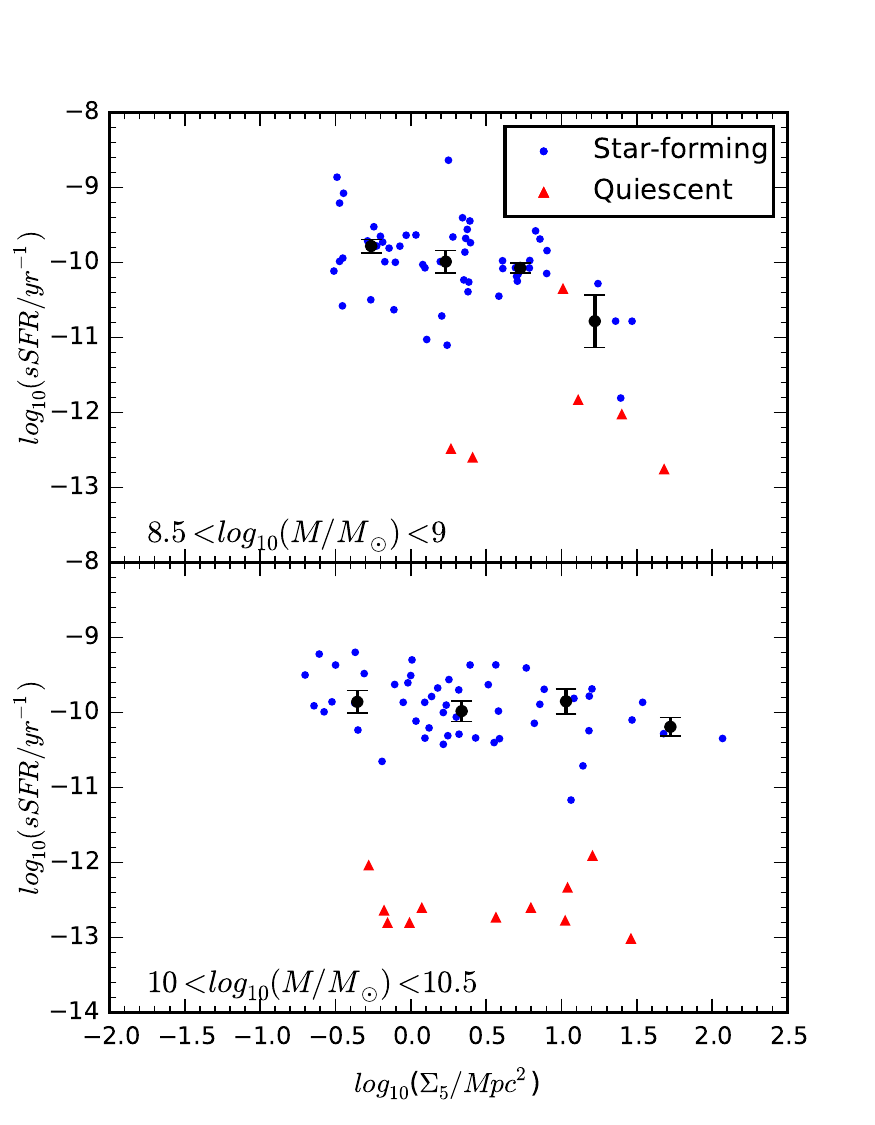}
\caption{Specific star formation rates as a function of $\Sigma_{5}$ for star-forming (blue points) and quiescent (red triangles) galaxies in two bins of stellar mass. The upper panel shows $55$ star-forming and $6$ quiescent galaxies with stellar masses in the range $10^{8}<\rm{M}_{*}<10^{9}$ M$_{\odot}$. A weak correlation between the specific star formation rates and local densities exists for galaxies in this mass range. The lower panel shows $51$ star-forming and $11$ quiescent galaxies with $10^{10}<\rm{M}_{*}<10^{10.5}$ M$_{\odot}$. There is a weak correlation between the specific star formation rate and local density in this mass bin. For both panels the black points show the median specific star formation rate for the star-forming galaxies in equally spaced bins of $\log_{10}(\Sigma_{5})$, with error bars determined as the bootstrapped error on the median in each bin. }\label{sfr_env_low_high}
\end{figure}

To control for the effect of the mass of a galaxy on its star formation rate, we have performed this analysis in two bins of restricted stellar mass. The upper panel of Figure~\ref{sfr_env_low_high} is constrained to stellar masses in the range $10^{8}$ to $10^{9}$ M$_{\odot}$, while the lower panel shows galaxies between $10^{10}$ and $10^{10.5}$ M$_{\odot}$. These mass bins were chosen from the peaks in the stellar mass distribution of our sample shown in Figure~\ref{m_env_dist}. Within each range of stellar mass we show the estimated specific star formation rate as a function of $\Sigma_{5}$. We test for the dependence of the specific star formation rate in galaxies on environment density with a Spearman rank correlation test. Star-forming galaxies in the higher mass sub-sample have a correlation coefficient of $\rho=-0.32$ with a p-value of $p=0.022$. The correlation coefficient between $\log_{10}(\Sigma_{5})$ and $\log_{10}(sSFR)$ in the lower mass subsample is $\rho=-0.31$ with $p=0.021$.

The correlation within the star-forming galaxies in the low-mass subsample is consistent with the results of some studies, such as \cite{Rasmussen12} who find that star-forming galaxies in groups have their star formation suppressed on average by $40\%$ relative to the field and that the trends are more prominent in lower stellar mass galaxies. A direct comparison between the current work and these previous results is difficult given that the environment density metrics differ between the two samples, as does the definition of a star-forming system. A future paper using SAMI data will examine the effect of group environments on the star formation distribution in galaxies and will allow a more direct comparison to other galaxy group studies.

\subsection{Dust-Corrected H$\alpha$ Radial Profiles}
We have calculated the radial profiles of dust-corrected H$\alpha$ emission for \SFNUM star-forming galaxies in the SAMI Galaxy Survey. These do not include galaxies that have been classified as quiescent in Section \ref{integrated_sfr}. The profiles trace the star-formation rate surface density with radius. In contrast to the radial profiles of broadband images, which can often be well described by a S{\'e}rsic profile, the distribution of H$\alpha$ is difficult to parametrise. The existence of spiral arms and asymmetric or clumpy structures in the ionised gas make the description of the radial distribution of gas with a single functional form problematic. These radial profiles are displayed in Figure~\ref{Profile_Grid}, where we have separated the profiles based on their stellar mass and local environment density. There is significant variation in the shape of each radial profile, and the median profiles show some change with environment density in a given mass bin, tending to be steeper at higher densities. At constant environment density, the normalisation of the radial profiles changes with stellar mass in accordance with the known SFR-$\rm{M}_{*}$ relationship \citep{Brinchmann04}, albeit with significant scatter. There is also some evidence for a steepening of the radial profiles with environment density, especially above $\mathrm{M}_{*}=10^{10} \, \mathrm{M}_{\odot}$, where the gradient of the median profile changes by $0.58 \pm 0.29 \, dex \, r_{e}^{-1}$ over the full range of environment densities. The best-fit central star formation rate surface density for the median profiles shows no significant change over the range of environments in our sample.

\subsubsection{Normalised H$\alpha$ profiles}
\begin{figure*}
\includegraphics{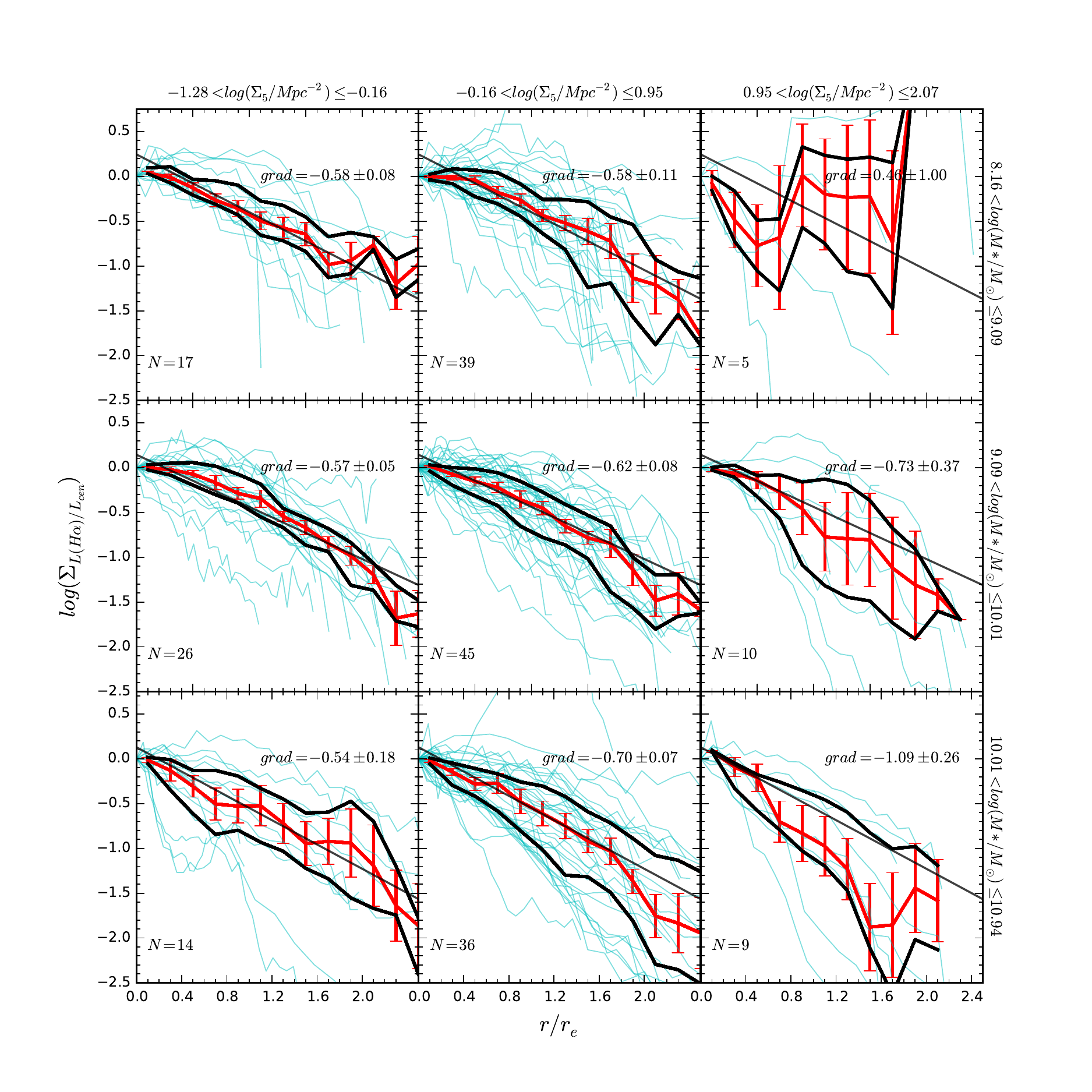}
\caption{Radial profiles for each of the star-forming galaxies in bins of stellar mass and environment density normalised at the centre. Each cyan coloured profile is the radial distribution of star formation divided by the value at the centre of the galaxy. In red we display the median normalised profile and the black lines are the $25^{th}$ and $75^{th}$ percentile profiles. Each row of panels contains galaxies within the same stellar mass bin, and mass increases downwards. Each column contains galaxies drawn from the same bin of environment density, with environment density increasing towards the right. As in Figure \ref{Profile_Grid}, these bins cover one third of the total range in $\log_{10}(\mathrm{M}_{*}/\mathrm{M}_{\odot})$ and $\log_{10}(\Sigma_{5}/\mathrm{Mpc}^{2})$. The diagonal black lines are the exponential fit to the median profiles of the central environment density bin for each row. In the bottom left of each panel we show the number of galaxies in each bin, while in the top right we show the gradient of an exponential fit to the median profile, with a bootstrapped $1\sigma$ error.}\label{Normalised_profile_grid}
\end{figure*}

Within the relatively narrow bins of stellar mass and environment density in Figure \ref{Profile_Grid} the normalisation of the star formation rate radial profiles vary by over a factor of one hundred. This is because of the stochastic nature of star formation and morphological variations between galaxies, or other factors that may not be attributable to the environment. These variations inhibit our ability to discern systematic differences in the star formation rate gradients across different environments. We correct for the effect of the normalisation of the star formation radial profiles by dividing each profile by its central value. In Figure \ref{Normalised_profile_grid} we show each normalised radial profile in the star-forming sample in three equally spaced, logarithmic bins of mass and environment density as in Figure \ref{Profile_Grid}. For each subset of galaxies we display the median profile and the $25^{th}$ and $75^{th}$ percentile profiles. The median profiles are approximately exponential, and we calculate slope parameters of the exponential fits in the same way as was done for Figure \ref{Profile_Grid}. These parameters are shown in Figure \ref{Normalised_profile_grid}.

In the lowest mass bin there is no evidence for a change in the gradient between the lowest and highest density environments, though we note that the small numbers in these bins do not allow strong conclusions to be made. For galaxies with $10  < \log{10}(\mathrm{M_{*}}/\mathrm{M_{\odot}}) < 11$ the median gradient of the normalised star-formation rate surface density radial profiles steepens from $0.54 \pm 0.18 \, dex \, r_{e}^{-1}$ in the lowest density environments to $1.09 \pm 0.26 \, dex \,r_{e}^{-1}$ in the highest density environments. The $0.55 \pm 0.32 \, dex \, r_{e}^{-1}$ increase in the gradient is significant at the $1.7 \sigma$ level.

\subsection{A Non-parametric Gradient Estimate: H$\alpha$ r$_{50}$ vs. continuum r$_{50}$} 

\begin{figure}
\includegraphics{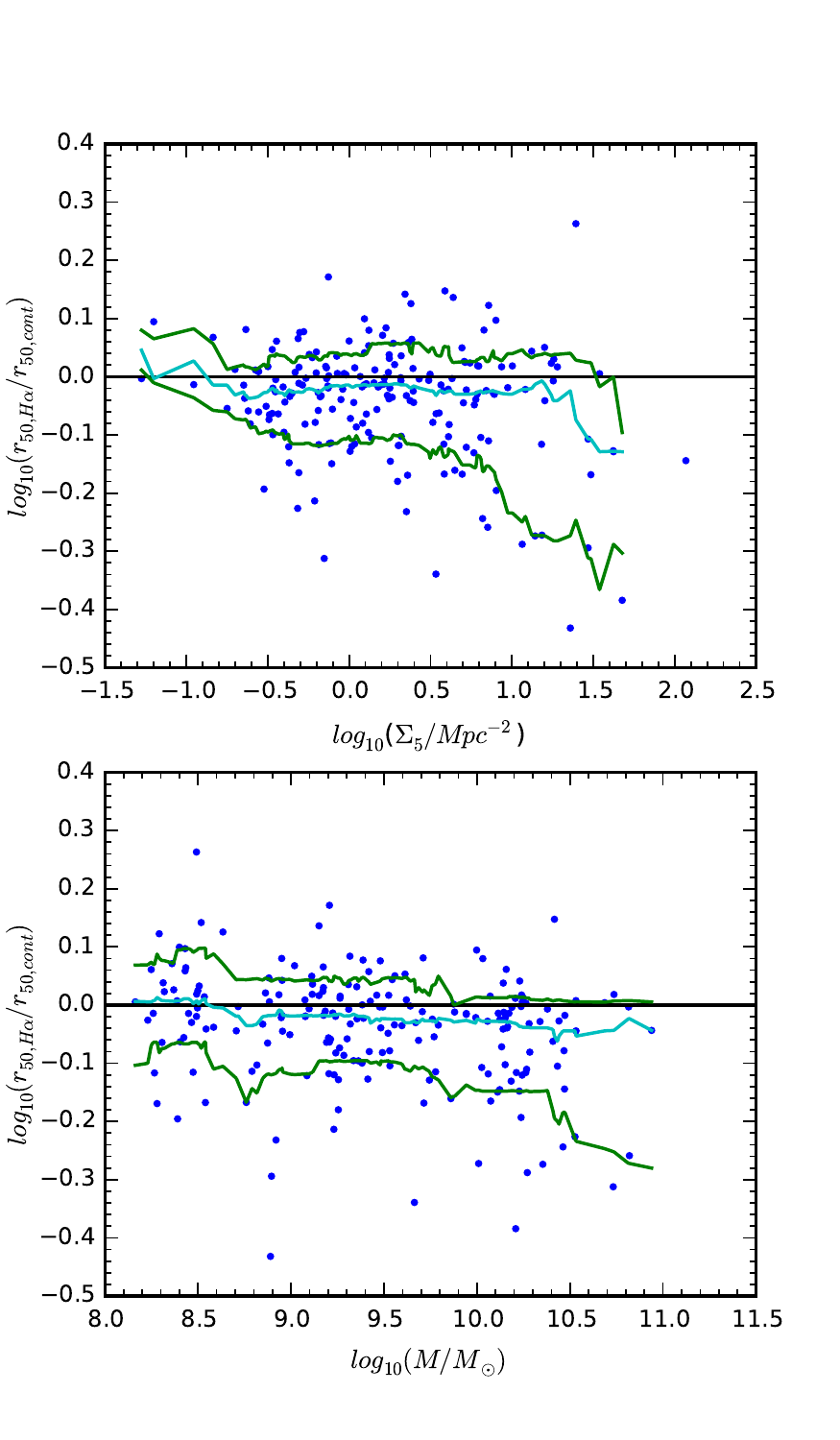}
\caption{\textit{Upper panel:} The scale radius ratio $r_{50,H\alpha}/r_{50,cont}$ for the star-forming galaxies in the sample as a function of the fifth-nearest neighbour local density. Higher values indicate more spatially-extended star formation while lower values are the signature of centrally-concentrated star formation. We show the $15^{th}$ (green), $50^{th}$ (blue) and $85^{th}$ (green) percentiles of  $r_{50,\mathrm{H}\alpha}/r_{50,cont}$ calculated in a sliding bin of width $0.5$ dex. While the $85^{th}$ percentile remains flat over the range of environments considered, the $15th$ percentile shows a sharp drop above $\log_{10}(\Sigma_{5}/Mpc^{2})=0.75$. \textit{Lower panel:} The scale radius ratio as a function of stellar mass for the same sample. The percentiles of the scale-radius ratio distribution show no strong trend with the stellar masses in this sample.}\label{r50ratios}
\end{figure}

The complexity of galaxy star formation morphology means that no single parametrisation will be sufficient to describe a galaxy's star-forming properties completely. We therefore turn to non-parametric methods of tracing the spatial signatures of star formation suppression in high-density environments. Non-parametric measurements of the star formation distribution in galaxies relative to the stellar light have been performed by a number of authors \citep[e.g.][]{Koopmann06,Cortese2012,Bretherton2013}. Following these previous studies, we can examine the possibility of outside-in quenching by measuring how concentrated current star formation is compared to previous episodes of star formation. We do this by comparing the half-light radius, r$_{50}$, of H$\alpha$ to that of the continuum as seen by SAMI. This analysis is performed by summing light within concentric elliptical annuli determined from the GAMA photometry to obtain curves of growth for extinction-corrected H$\alpha$ and red continuum light within the SAMI aperture. Some examples of dust-corrected H$\alpha$ curves of growth are displayed in the right hand column of Figure \ref{rpeg} in black. The radius at which these curves of growth reach 50\% of their maximum is defined to be r$_{50}$. In Figure \ref{r50ratios} we show the ratio $r_{50,H\alpha}/r_{50,cont}$ as a function of the local surface density and stellar mass. If $r_{50,H\alpha}/r_{50,cont} \gtrsim 1$, star formation is spatially extended and the current buildup of stellar mass in the galaxy is occurring in the outer parts. If $r_{50,H\alpha}/r_{50,cont} < 1$ then star formation is centrally concentrated and stellar mass in the galaxy is now predominantly accumulating in the inner region. This measurement provides a differential test to examine the relative spatial extent of star formation over a range of masses and environments. A discussion of some possible issues with this measurement can be found below in Section \ref{R50Biases}. In the star-forming sample $\log_{10}(r_{50,H\alpha}/r_{50,cont})$ has a median of $-0.020$ and a standard deviation of $0.099$. The Spearman rank correlation coefficient between $\Sigma_{5}$ and $r_{50,H\alpha}/r_{50,cont}$ is $\rho=-0.10$ with $p=0.14$. 

Figure \ref{r50ratios} shows that galaxies in higher density environments have a greater probability of having centrally concentrated star-formation than do galaxies in low-density environments. It is important to note that dense environments in this sample harbour both galaxies with extended star-formation, as seen in the low-density regions, as well as galaxies with a more compact star formation morphology. However, the distribution of scale-radius ratios changes with increasing environment density. In Figure \ref{r50ratios} we see that the scatter in the distribution of $r_{50,\mathrm{H}\alpha}/r_{50,cont}$ increases significantly with environment density, and that this scatter is biased towards galaxies with more centrally concentrated star formation in dense environments.

\begin{figure}
\includegraphics{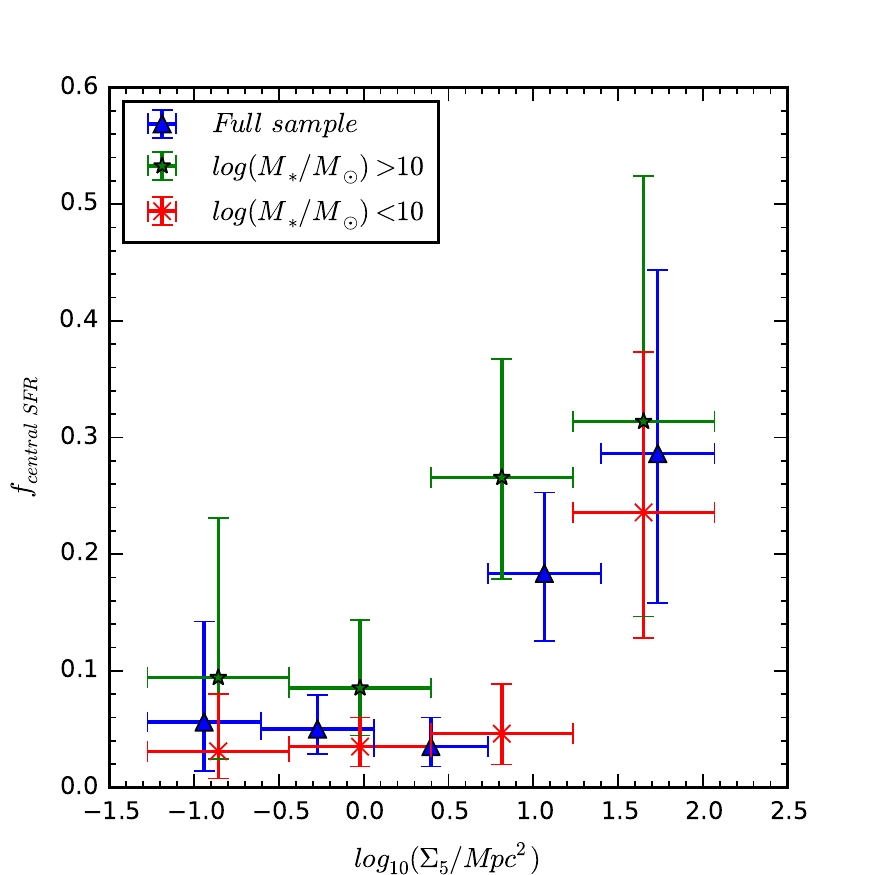}
\caption{Fraction of galaxies with centrally concentrated star-formation as a function of local environment density. Blue triangles show the fractions for the full star-forming sample, red crosses are the fractions for star-forming galaxies with $log_{10}(\mathrm{M}_{*}/\mathrm{M}_{\odot}) < 10$, while green star symbols are the fractions for galaxies with $log_{10}(\mathrm{M}_{*}/\mathrm{M}_{\odot}) > 10$,. Fractions and vertical errors are calculated after \protect\cite{Cameron11} while the horizontal error bars show the range in environment densities over which each fraction was computed.}\label{cen_con_frac_env}
\end{figure}

We define galaxies with $\log_{10}(r_{50,H\alpha}/r_{50,cont})<-0.2$ (i.e. further than two standard deviations below the median) as `centrally concentrated'. The fraction of galaxies with centrally concentrated star formation, increases significantly with environment. We see in Figure \ref{cen_con_frac_env} that below $\Sigma_{5}=10^{0.5} \, \mathrm{Mpc^{-2}}$ typically only $5 \pm 4\%$ of galaxies show this centrally concentrated star forming morphology. In higher density environments the fraction of centrally-concentrated star-forming galaxies rises to $30 \pm 15 \%$.

In Figure \ref{cen_con_frac_env} the red and green points represent the fractions of star-forming galaxies with centrally-concentrated star formation for stellar masses below and above stellar masses of $10^{10} \, \mathrm{M}_{\odot}$ respectively. There is some evidence that galaxies with stellar masses greater than $10^{10} \, \mathrm{M_{\odot}}$ show centrally concentrated star formation more readily than do galaxies with stellar masses below $10^{10} \, \mathrm{M_{\odot}}$. The fraction of galaxies with centrally concentrated star formation is the same for both the high and low stellar mass subsamples over all environments except for in the second highest environment density bin. Here the high mass subsample shows a higher fraction of centrally concentrated star formation than the low mass subsample. This may imply that higher mass galaxies are more susceptible to environmental effects than lower mass galaxies. This seems unlikely given that the efficiency of environmental quenching mechanisms such as ram pressure stripping and tidal disruption of the star-forming disc is predicted to be reduced in higher mass galaxies. An alternative explanation for this may be that any process that causes star formation to be centrally concentrated occurs on much shorter timescales in lower mass galaxies or that low mass galaxies are quenched in a qualitatively different manner in all but the most dense environments.

\subsubsection{Systematic biases and sources of error for the scale-radius ratio}\label{R50Biases}
Errors on the scale-radius ratio for each galaxy are extremely difficult to quantify. Small uncertainties in this quantity arise from random errors on the measured flux and photometric fits to the galaxy's ellipticity and position angle. These error terms are small in comparison to the uncertainties induced by observational effects.

Beam-smearing, which will force the scale-radius ratio towards $1$, and aperture effects will have a larger contribution to the systematic biases in this measurement. We are only able to measure the scale-radius ratio from the parts of the galaxy which fall within the $15\arcsec$ SAMI hexabundle. Given that the distribution of effective radii for the galaxies in our sample (see Figure \ref{AllParams}) includes a number of galaxies which are large relative to the SAMI aperture, it is possible that aperture effects may introduce biases in our measurement of the scale-radius ratio. This bias is dependent on the distribution of star formation in each galaxy observed. If the star formation in a galaxy is radially extended, then the scale-radius ratio measurement will be lower than the true value. For systems with more centrally concentrated star formation, the scale-radius ratio measured within the IFU aperture will be higher than the true value. If the radius out to which we observe the galaxies in the sample was randomly distributed across all environments, these aperture effects would add scatter to the $r_{50,H\alpha}/r_{50,cont}$ vs. $\Sigma_{5}$ relation, but would not induce a trend. Indeed, when we look for a correlation between $r_{50,H\alpha}/r_{50,cont}$ and $7.5/r_{e}$ (the effective radius coverage of the SAMI aperture), we find no significant correlation, with Spearman's $r=-0.03$ with a p-value of $0.61$. When we control for the effects of galaxy mass and environment on the sample with a partial correlation analysis, this correlation coefficient is reduced in magnitude to $-0.02$ with a p-value of $0.71$. Moreover, when the effect of radial coverage on the correlation between $r_{50,H\alpha}/r_{50,cont}$ vs. $\Sigma_{5}$ is controlled for, we see no reduction in its strength. We also see no correlation between $\Sigma_5$ and the radial coverage of galaxies within our sample. There is therefore no trend between $\Sigma_{5}$ and $r_{50,H\alpha}/r_{50,cont}$ imposed by our sample selection. We conclude that the effective radius coverage does not affect our ability to discern environmental trends, though it may affect the measured value of the ratio for individual galaxies. Since the dominant error terms for individual galaxies are unquantifiable with the current data, we do not include them in the figures.

\subsubsection{Scale-radius ratio and the D$_{n}4000$ gradient}\label{r_50 and D4000 gradients}

\begin{figure}
\includegraphics{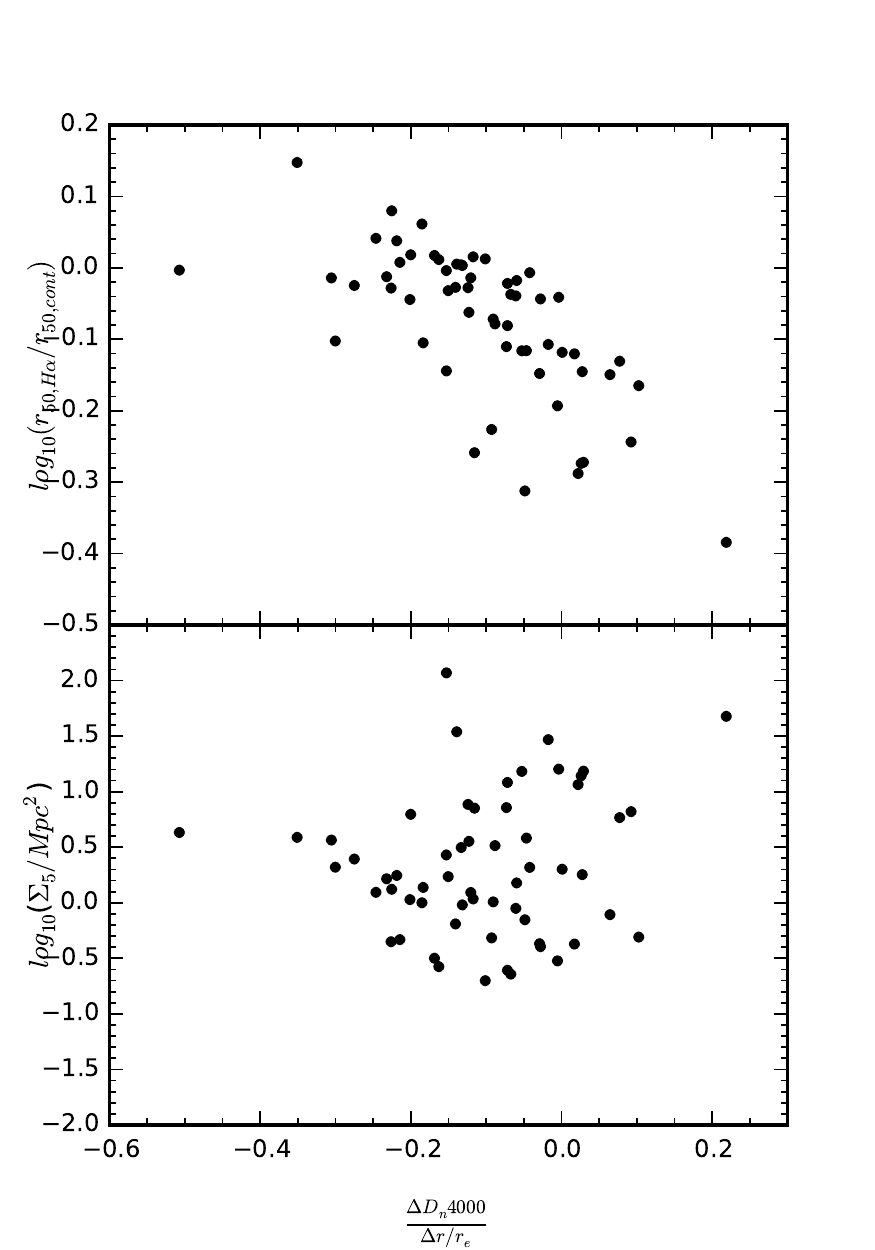}

\caption{Relationship between $\Sigma_{5}$, $r_{50,H\alpha}/r_{50,cont}$ and the D$_{n}4000$ gradient for the $60$ star-forming galaxies with $\rm{M}_{*}>10^{10}$ M$_{\odot}$. The upper panel illustrates the correlation between the scale-radius ratio and the D$_{n}4000$ gradient. Galaxies with positive D$_{n}4000$ gradients have older stellar populations towards their edges and tend to have a centrally concentrated H$\alpha$ distribution. The lower panel shows no significant correlation between the D$_{n}4000$ gradient and $\log_{10}(\Sigma_{5}/Mpc^{2})$.}\label{r50_d4000}
\end{figure}

The lifetimes of \ion{H}{ii} regions are short relative to the evolutionary timescale of galaxies. To test for the possibility of short-timescale variability in the H$\alpha$ distribution, we consider the scale radius ratio, $r_{50,H\alpha}/r_{50,cont}$ in conjunction with the radial gradient in the D$_{n}4000$ strength. This gradient has the additional advantage of being able to approximate age gradients in the stellar populations of these galaxies. 
In Figure~\ref{r50_d4000} we present the relationship between the D$_{n}4000$ gradient, $r_{50,H\alpha}/r_{50,cont}$ and $\Sigma_{5}$ for galaxies with stellar mass exceeding $10^{10}$ M$_{\odot}$. This mass range is chosen because the continuum in these galaxies has sufficient S/N to make an accurate estimate of the D$_{n}4000$ break out to large radius. 

\begin{figure*}
\includegraphics[trim=0.4cm 0cm 0.4cm 0cm, clip,scale=0.8]{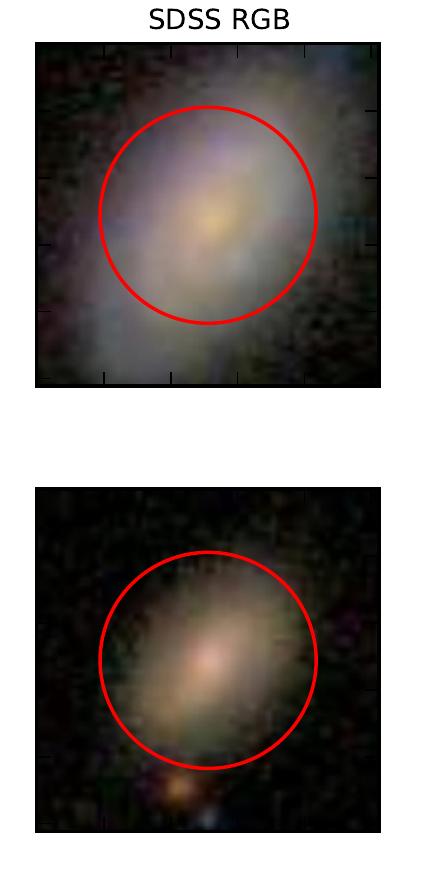} \includegraphics[trim=0.0cm 0cm 0.0cm 0cm, clip,scale=0.8]{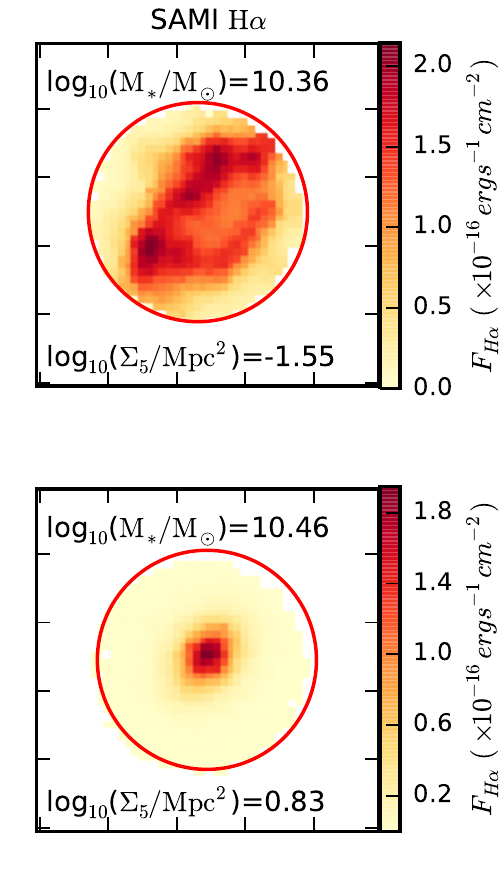}  \includegraphics[trim=0.0cm 0cm 0.0cm 0cm, clip,scale=0.8]{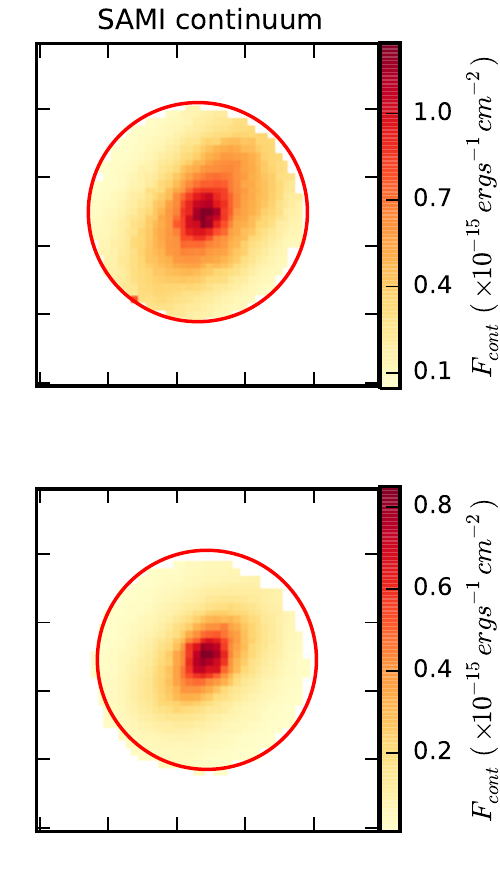}  \includegraphics[trim=0.4cm 0cm 0.0cm 0cm, clip,scale=0.8]{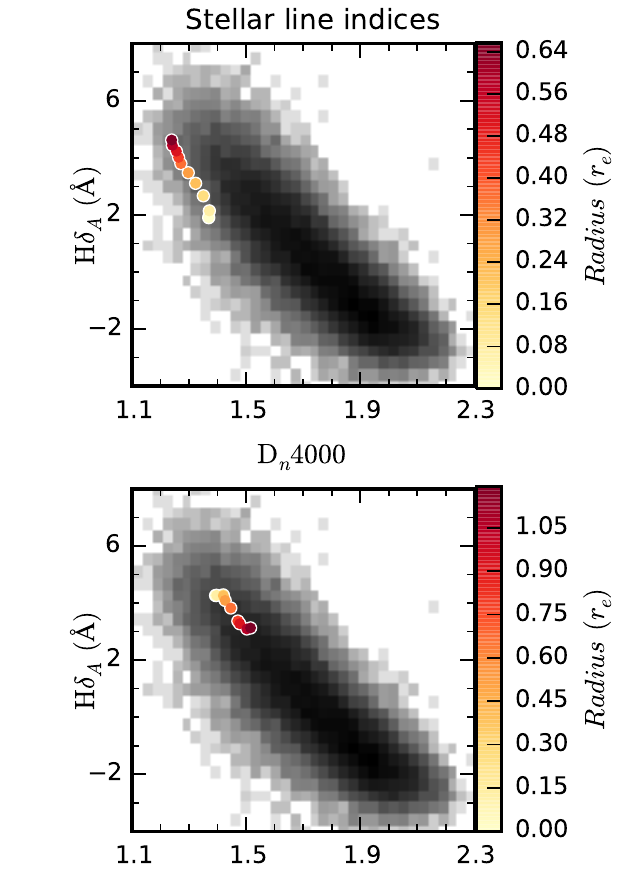}
\caption{Examples of the relationship between the star formation morphology of a galaxy and the inferred age gradients in the underlying stellar populations. Each row shows (from left to right) the SDSS image, the SAMI H$\alpha$ map, the SAMI continuum map and the radial distribution of the D$_{n}4000$ strength and the emission-subtracted H$\delta$ absorption equivalent width index. The red circles in the three left columns are $15 \arcsec$ in diameter, encircling the SAMI field of view, and each box is $25\arcsec$ in width and height. In the final column, the greyscale background traces the distribution of D$_{n}4000$ and H$\delta_{A}$ for a large sample of galaxies from the SDSS \citep{Kauffmann03c}. The coloured points are measured values from the SAMI data cube, with yellow points extracted from the centre of the galaxy and red points from towards the edge. The upper row shows a galaxy (GAMA 41144) with an extended H$\alpha$ morphology and an older stellar population in the centre, as evidenced by the higher D$_{n}4000$ and lower H$\delta_{A}$. The lower row presents a system (GAMA 492384) from a more dense environment with a concentrated star formation morphology and older stellar populations towards the edge of the galaxy disc.}\label{D4000_examples}
\end{figure*}

We find that the D$_{n}4000$ gradient is anti-correlated with the scale-radius ratio. This implies that galaxies that are identified as being quenched in their outskirts by the scale-radius ratio also have older stellar populations toward their edges relative to those which show no evidence for H$\alpha$ truncation. Galaxies with more radially extended star formation will have a higher proportion of young stars in their outskirts, and therefore exhibit younger stellar populations towards their edge. The observed relationship corroborates the evidence from the scale-radius ratios and H$\alpha$ luminosity surface density profiles that increasing environment density reduces the star formation in the outskirts of galaxies. This correlation is illustrated in Figure \ref{D4000_examples}, where we see two galaxies from different density environments with different H$\alpha$ morphologies. The age gradients inferred from the D$_{n}4000$ and H$\delta_{A}$ measurements are opposite for these two systems.

\subsubsection{Scale-radius ratio and the specific star formation rate}\label{r50_ssfr_section}
\begin{figure}
\includegraphics{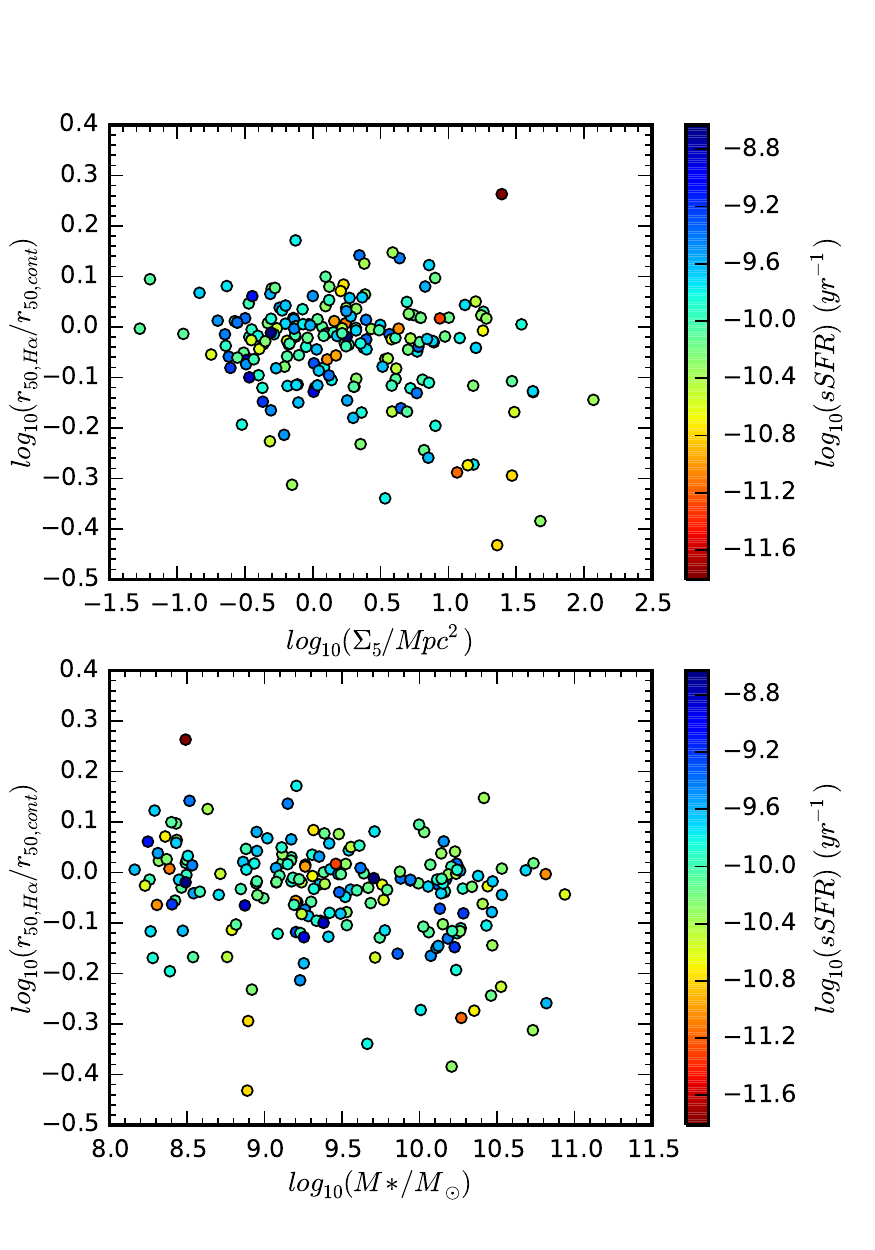}
\caption{Scale radius ratio $r_{50,H\alpha}/r_{50,cont}$ as a function of $\log_{10}(\Sigma_{5}/Mpc^{2})$ and $\log_{10}(\rm{M}_{*}/\rm{M}_{\odot})$ for \SFNUM\ star-forming galaxies across the full range of masses. Each point is coloured by the specific star formation rate in that galaxy. Galaxies in the densest environments with the lowest specific star formation tend to have the most centrally concentrated star formation. Conversely, galaxies in lower density environments with low specific star formation rates have a radially extended star formation morphology, implying that different quenching mechanisms may dominate in different density environments.
The lower panel shows how the specific star formation rate varies with mass and the scale-radius ratio. There is no strong correlation between stellar mass and specific star formation rate for the star-forming galaxies, though more massive galaxies with centrally concentrated star formation do tend to have a lower specific star formation.}\label{sSFR_r50}

\end{figure}

We adopt the definition of quenching that is the movement of a galaxy off the star formation main sequence to a lower specific star formation rate. By examining the spatial distribution of current star formation in galaxies as they make this transition we can better understand the mechanisms responsible for this evolution. While the current sample is the largest studied to date, it is still of insufficient size to make strong statements about the statistics of particular quenching mechanisms in different environments. We note that systems with specific star formation rates that are lower than the star-forming main sequence are located in a range of environments and have a variety of H$\alpha$ morphologies. The upper panel of Figure \ref{sSFR_r50} shows how the specific star formation rates of galaxies depend on their local environment density and the relative extent of the current star formation. 
The scale radius ratios of galaxies with below average specific star formation rate (i.e. $\mathrm{sSFR}<10^{-10} \, \mathrm{yr}^{-1}$) in environments above and below $\log_{10}(\Sigma_{5}/\mathrm{Mpc})=0.5$ show a significant difference. Below this environment density the mean log scale-radius ratio in galaxies is $-0.01 \pm 0.01$, but above this density, the mean log scale-radius ratio $-0.07 \pm 0.02$. That is, galaxies in lower density environments that sit below the star formation main sequence exhibit a more diffuse and extended star formation morphology. Galaxies in higher density environments with star formation rates lower than the star formation main sequence have a more centrally concentrated star formation morphology. The dominant mechanism responsible for quenching the star formation in galaxies in low-density environments seems to differ qualitatively from the mechanisms reducing the star formation in higher density environments.
 
In the lower panel of Figure \ref{sSFR_r50} we see the relationship between the scale-radius ratio, the stellar mass and the specific star formation rate for galaxies in our sample. There is no significant correlation between the stellar mass of a star-forming galaxy and the relative spatial extent of its star formation. Furthermore, galaxies with specific star formation rates below $10^{-10.8} \, yr^{-1}$ do not occupy any particular region of this parameter space and seem to have no tendency to have either their star-formation either spatially-extended or centrally-concentrated.

\section{Discussion}\label{Discussion}
We have used SAMI integral field spectroscopy to examine variations in the radial distribution of star formation in galaxies as a function of their local environment density. Our sample of \SFNUM star-forming galaxies covers nearly three orders of magnitude in both stellar mass and local environment density. Using the dust-corrected H$\alpha$ maps of the galaxies we have computed radial profiles of current star formation surface density. 

Our analysis shows a change in the median radial distribution of current star formation between high and low-density environments for galaxies with $\log_{10}(\rm{M}_{*}/\rm{M}_{\odot})$ $>9$. While this was not seen by \cite{Brough13} this is unsurprising since the large variation in star formation morphologies in galaxies at a given mass and environment density makes environmental trends difficult to observe in small samples. The current sample is over an order of magnitude larger, allowing us to average over a wide range of galaxy morphologies that would dominate the conclusions drawn from small samples.

The radial profiles of star formation surface density in star-forming galaxies show a systematic change in normalisation with stellar mass. Higher mass galaxies have a larger star-formation rate surface density at all radii than do their low-mass counterparts. This is consistent with the known relationship between the total star-formation and stellar mass of a galaxy. In Figure \ref{Profile_Grid} we do not see a decline in the central star formation rate surface density with environment density for galaxies of a given mass. However, the global specific star-formation rates shown in Figure \ref{ssfr_m_env_frac} do show a decline. The apparent disagreement between these two measurements is reconciled by the steeper star formation rate surface density gradients in higher density environments. This implies that the reduction in star-formation in galaxies in high density environments occurs primarily in their outskirts. This environmental effect may be responsible for the difference between the results of, for example, \cite{vonderLinden2010} and \cite{Wijesinghe12}. As \cite{Wijesinghe12} used H$\alpha$ equivalent widths derived from single fibre spectroscopy at the centre of the galaxy as a proxy for specific star formation rate, and calculated the star formation rate assuming that the emission line intensity as measured within the fibre was representative of the entire galaxy \citep{Wijesinghe2011}. If environmental effects primarily influence the outskirts of galaxies then the integrated star formation rates in dense environments may have been systematically overestimated \citep[see][for a discussion of these aperture effects]{Richards16}. Thus, the scenario of short timescale quenching that was supported by the results of \cite{Wijesinghe12} is not favoured by our data.

A simple parametrisation of the radial distributions of H$\alpha$ luminosity is insufficient to diagnose the mechanisms for quenching. Variation in the structure of the galaxies, for example the presence of spiral arms, bars and other morphological features, as well as inhomogeneities in the galaxy angular size and observing conditions means that comparing star formation rate gradients directly is difficult. While the median profiles of H$\alpha$ luminosity surface density roughly follow an exponential profile we lose information about the distribution of galaxy properties including the underlying stellar distribution with this type of comparison. Instead the spatial distribution of ongoing star formation can be quantified relative to that of previous star formation. The non-parametric scale radius ratio provides a differential test that is more robust to variation between galaxies as it is not reliant on an invalid parametrisation of a complicated structure. 
We find that the scale radius of H$\alpha$ emission in galaxies in high-density environments is reduced relative to the scale radius of previous star formation by approximately 20\% over the range of environment densities considered here. The reduction in the scale-radius ratio in increasingly dense environments is consistent with the environmental dependence of the median radial profiles seen in Figure \ref{Profile_Grid} and the normalised profiles in Figure \ref{Normalised_profile_grid}. The fraction of galaxies which show a centrally concentrated star formation morphology increases from $\sim 5 \%$ below $\log_{10}(\Sigma_{5}/\rm{Mpc^2})=0.5$, the density above which the passive fraction of galaxies increases, to $29\%$ above this density threshold.

Moreover, we have observed that in galaxies with a more centrally concentrated H$\alpha$ distribution, the gradient of the D$_{n}4000$ spectral feature is shallower. This is indicative of an older stellar population in the outskirts relative to galaxies with ongoing star formation in their peripheries. In the most extreme cases, the D$_{n}4000$ spectral feature is strongest towards the edge of the galaxy, indicative that all significant star formation in their edges has ceased. Due to the degeneracy between age and metallicity in setting the strength of D$_{n}4000$, it is not possible to calculate the age gradient from this measurement alone. Even so, the timescale over which D$_{n}4000$ is likely to change is longer than the typical lifetimes of HII regions. The fact that the correlation between the D$_{n}4000$ gradient and $r_{50,H\alpha}/r_{50,cont}$ is so strong implies that the suppression of star formation in the outskirts of galaxies is maintained on timescales longer than the lifetimes of individual HII regions.

This observed outside-in mode of quenching is qualitatively different to the findings of \cite{Welikala2008}, who found that local environment tended to reduce the star-formation rate in the central regions of galaxies. The disparity between these results and our own highlights the difficulties associated with spectral energy distribution fitting from broadband imaging given the degeneracies between dust-extinction, stellar age and metallicity in galaxy colours. The ability of integral field spectroscopy to obtain independent measurements of the dust extinction in galaxies will be invaluable to future studies of environment quenching.
\cite{Peng2010} observed a tendency for galaxies of larger mass to quench more readily than galaxies of lower mass, independent of their local environment. While there is some evidence for a correlation between galaxy stellar mass and $r_{50,H\alpha}/r_{50,cont}$ in Figure \ref{r50ratios}, this trend is very weak. Moreover, the measured steepening of the star-formation rate gradients in Figures \ref{Profile_Grid} and \ref{Normalised_profile_grid} is greatest for the most massive galaxies in the sample. This does not necessarily indicate that galaxies of different masses experience different quenching mechanisms or that these processes are stronger for the most massive galaxies. Instead it is likely that the timescale over which quenching occurs is related to the galaxy mass. If lower mass galaxies are quenched rapidly by any means then observing these systems as quenching proceeds is less likely.

\subsection{Environment-enhanced star formation}
While we have only considered the quenching of star formation by environmental effects here, there is a growing body of evidence that suggests that 
star formation can be enhanced in group environments \citep[e.g.][]{Kannappan13, Robotham14}. This enhancement can be from either gas-rich mergers, tidal torques from galaxy interactions or from the accretion of cold or hot gas from the intergalactic medium. \cite{Sancisi08} argue that a large fraction of this gas must be accreted directly from the intergalactic medium, rather than through gas-rich mergers. Simulations by \cite{Keres05} showed that cold-mode accretion does not dominate in galaxy groups with space densities above $\sim 1 \,\rm{Mpc}^{-3}$ (using a modified $\Sigma_{10}$ density measure), implying that hot-mode accretion must re-supply the gas in galaxies in dense environments.

Observationally, the infall of cold gas onto a low mass ($\rm{M}_{*}<10^9 \, \mathrm{M}_{\odot}$) system will be manifest in localised, low metallicity and offset star-forming regions \citep[e.g.][]{Richards14,SanchezAlmeida2015,Ceverino15}. Given that the unusual galaxy studied by \cite{Richards14} is a part of our sample ($\mathrm{M}_{*}=10^{8.5} \, \mathrm{M}_{\odot}$), this effect is certainly present in our data and helps explain the scatter in $r_{50,H\alpha}/r_{50,cont}$ at low stellar mass. At higher stellar masses the effects of gas accretion on star formation morphology are more difficult to diagnose as the specific morphological signatures can vary depending on the details of the accretion process. A better way of classifying the accretion of gas onto galaxies with $\rm{M}_{*}>10^{10} \, \rm{M}_{\odot}$ would be to investigate their gas kinematics and metallicity, though this analysis is beyond the scope of this paper (but see Bryant et al. \emph{in prep.}).

\subsection{The Mechanism of Environmental Quenching}
The suppression of star formation towards the edges of galaxies in dense environments is consistent with a number of proposed quenching mechanisms. Ram pressure stripping is commonly cited as a mechanism to remove gas from the edges of discs in group \citep{Rasmussen06} and cluster galaxies \citep{Koopmann04a,Koopmann04b}. The removal of cold gas from a galaxy by this method is possible only if the velocity of the galaxy through the intergalactic medium is such that the ram pressure force exerted on the interstellar medium is sufficient to overcome the gravitational potential of the galaxy. For the majority of more massive galaxies in our sample this is probably not the case, given that they are drawn from field environments or groups with a velocity dispersion typically below 300 km s$^{-1}$. \cite{Nulsen1982} points out that under these circumstances ram pressure is unlikely to be the dominant method of gas removal from galaxies, and instead suggests that turbulent or laminar viscous stripping may dominate. The removal of gas by viscous stripping in elliptical cluster galaxies has been modelled by \cite{Roediger14}, who show that this method acts in an outside-in way that is qualitatively consistent with what has been observed here, though the presence of a cool X-ray tail in the models cannot be verified for SAMI galaxies at present. The effect of viscous stripping is likely to be more efficient in galaxy clusters than in groups and the field. This will be the subject of future work on the SAMI Galaxy Survey cluster sample.

Tidal interactions are another mechanism that has been proposed to be responsible for a significant fraction of the passive population of galaxies in dense environments. Simulations \citep{Hernquist89,Moreno2015} have shown that tidal disturbances generated by a near miss with another galaxy will produce disc instabilities, which drive gas towards the galaxy's centre. The resulting enhancement of star-formation is predicted to persist on timescales of proximately $1 \, \mathrm{Gyr}$. Since this mechanism is most sensitive to the nearest neighbour distance, the correlation between tidal suppression of star formation and $\Sigma_{5}$ could potentially be weak.

It is likely that all of these quenching mechanisms act on galaxies in dense environments to some degree. \cite{Yonzin2015} suggested that within galaxy groups the star formation distribution in a galaxy is influenced simultaneously by both tidal interactions causing central enhancement of star-formation and ram-pressure stripping reducing the star-formation rate at large galactic radius. Several profiles in the highest density environment bins in Figure \ref{Profile_Grid} do seem to show evidence of both central enhancement of star-formation and its reduction at large radius, though examples of such systems are not numerous enough to alter the median profiles.

The SAMI observations are evidence for different modes of quenching occurring in different environments. In Section \ref{r50_ssfr_section} we showed that for $\log_{10}(\Sigma_{5}/\mathrm{Mpc^{2}}) > 0.5$, galaxies with below average specific star formation rates tend  to have their star formation concentrated towards their centres. In environments of lower density, systems with low specific star formation rates have more extended star formation. The spatial dependence for the quenching of star formation in galaxies varies across the range of environmental densities currently sampled by our survey. 

Further insight into the nature of environmental effects on star formation in galaxies will be obtained by investigating how the spatial distribution of star formation is affected by the galaxy group properties. The separation of galaxies in dense environments into satellite and central systems has featured heavily in recent studies of environment processes \citep[e.g.][]{vandenBosch08,Peng2012}. This and the increased sample size that will be afforded by the full SAMI galaxy survey will enable us to investigate these aspects of environmental star formation quenching in much more detail. A study of the effects of galaxy group properties on star formation will be presented in a future paper.

\subsection{How many galaxies do we need to observe?}
While the number of systems that we have studied is a factor of $10$ larger than that of \cite{Brough13}, the environmental quenching trends when separated into bins of mass, are too weak to quantify for the whole range of galaxy masses with the current sample. However, the results obtained from the present study set the scene for future investigations of this topic.

The SAMI Galaxy Survey sample is drawn from mass selected volumes derived from the GAMA survey. As a result our coverage of the whole range of environment densities is not uniform. The majority of galaxies in our sample exist in low and intermediate density environments where the intrinsic scatter in galaxy properties is dominant. We have circumvented this inhomogeneity in the environment density by splitting our sample and using metrics for comparison that are independent of the total number of galaxies in each bin, such as the fraction of galaxies with centrally concentrated star formation. These fractions still have an associated error, and in future studies of larger samples we will reduce the error bars and achieve a comparable statistical weight in a larger number of stellar mass and environment density bins.

If we assume that the observed trend between the fractions of galaxies with centrally concentrated star formation and $\Sigma_{5}$ holds, we can estimate the numbers of galaxies required to achieve $1\sigma$ errors of $\pm 5\%$ on the fractions. We do so by generating an ensemble of beta distributions defined by a variety of galaxy numbers in the same ratio as observed in our sample. For each beta distribution the $1\sigma$ error interval is define by the range between the $15.9^{th}$ and $84.1^{th}$ percentiles. In low-density environments where $\sim 95 \%$ of star forming galaxies have extended star formation morphologies we need only $\sim 25$ galaxies per bin in stellar mass and environment density to achieve the $5\%$ error bar. In high density environments, since the fraction of galaxies with centrally concentrated star formation is much higher, we need a larger number of galaxies in each bin. We estimate that approximately $75$ galaxies are required in high density environments to constrain the faction to better than $5\%$. Therefore, if we were to repeat the analysis from Figure \ref{cen_con_frac_env}, with three environment density bins showing low fractions of centrally concentrated star formation and two bins showing higher fractions of centrally concentrated star formation we would require $225$ galaxies per mass bin, with a distribution that is highly weighted towards the high density environments.

The full SAMI survey will include $\sim 3400$ galaxies. This will allow us to separate galaxies into finer bins of stellar mass and local environment density. If we reject galaxies at the same rate as we have for this analysis (as detailed in Section \ref{Sample_selection}; i.e. from an initial sample of $808$ galaxies down to \SFNUM), we will have a total of $900$ galaxies at our disposal for further analysis. Under the assumption that we will require $225$ galaxies to detect an environmental trend with $5\%$ errors, this implies that we will be able to study the environmental trends in roughly $4$ bins of galaxy stellar mass. With the completion of SAMI and other large-scale IFU surveys we will have an unprecedented view of the environmental processes occurring in galaxies.

\section{Conclusion}\label{Conclusion}

We have used SAMI integral-field spectroscopy to study the spatial distributions of ongoing star formation in a sample of \SFNUM\ star-forming galaxies as a function of their local environment densities. 

We have shown that the derived integrated H$\alpha$ star formation rates from integral field spectroscopy will be underestimated by approximately $9\%$ if a single average dust correction is applied to the galaxy. The non-linear nature of dust extinction demands that the attenuation must be accounted for locally within different regions of a galaxy. Failure to perform a local spaxel-by-spaxel dust extinction correction results in a systematic reduction of the total measured SFR of $\sim 8 \%$ in the most star-forming galaxies.

Analysis of the star formation rate surface density radial profiles of \SFNUM\ galaxies in our sample has shown a large variation in the radial distributions of star formation in our galaxies. For any given mass or environment density the normalisation of a galaxy's star formation rate radial profile can vary by more than a factor of $10$. For our star-forming sample we note that the normalisation of the star formation increases almost linearly with stellar mass in accordance with the known star-formation rate versus stellar mass relation. 
There is no significant relationship between the central star formation rate surface density and the local environment density, but we note a $2\sigma$ significance steeping of the H$\alpha$ radial profiles in high density environments in galaxies with masses in the range $9.92 < \log_{10}(\mathrm{M}_{*}/\mathrm{M}_{\odot}) < 10.94$. This is consistent with the relationship between the specific star formation rates and $\Sigma_{5}$ in Figure \ref{ssfr_m_env_frac}, and implies that the environmental suppression of star formation must occur first in the outskirts of galaxies

When the central star-formation rate surface density in a galaxy is controlled for, we see a significant steepening of the profiles in higher density environments. For galaxies in the highest mass bin ($9.92 < \log_{10}(\mathrm{M}_{*}/\mathrm{M}_{\odot}) < 10.94$) we observed the median normalised star formation rate profiles to steepen from $-0.54 \pm 0.18 \, \mathrm{dex} \, r_{e}^{-1}$ to $-1.09 \pm 0.26 \, \mathrm{dex} \, r_{e}^{-1}$.

We have also shown that $(30 \pm 15)\%$ of galaxies in high-density [$\log_{10}(\Sigma_{5}/\rm{Mpc}^{2})>0.5)$] environments exhibit a centrally concentrated star formation distribution. In low-density environments only $(5 \pm 4)\%$ show similar star formation morphologies. This is taken as evidence that the occurrence of outside-in quenching is more common in dense environments and is consistent with the findings of studies in clusters \citep[e.g.][]{Koopmann04b,Koopmann06}.
This conclusion is supported by the observed radial gradients in the D$_{n}4000$ flux ratio. Galaxies which exhibit the most extremely centrally concentrated H$\alpha$ also have stronger D$_{n}4000$ towards their edges, in contrast to systems which show extended H$\alpha$ morphologies. As the D$_{n} 4000$ spectral feature is sensitive to star formation quenching on timescales of up to $\sim1$ Gyr, we estimate that this outside-in quenching must occur faster than this, though careful modelling and a more comprehensive study of the spectral features will be required to determine the timescales accurately.

The observed correlations between the star formation morphology of galaxies and the local environment density are significant but weak. These trends appear to be dominated by the intrinsic variation in galaxy properties across all environments. For this reasons sample sizes of over $225$ galaxies per mass bin will be required to ensure the statistical significance of the results of future studies.

In lower density environments ($\log_{10}(\Sigma_{5}/Mpc^{2}) < 0.5$), the smaller fraction of passive systems indicates that quenching must be occurring at a lower rate than in higher density regions. In these environments galaxies with the lowest specific star formation rates appear to have spatially extended H$\alpha$ morphologies which we take to be the qualitative signature of different mechanisms (such as starvation) acting to suppress the star formation in galaxies.

Galaxies of higher stellar mass are more likely to show the signatures of outside-in quenching. Even when the effect of correlation between mass and environment density is controlled for, this relationship persists. We must therefore conclude one of several things: 1) that higher mass galaxies are more susceptible to environmental effects than low mass galaxies, 2) the environmental mechanisms that quench high mass and low mass galaxies are qualitatively different, or 3) that the outside-in quenching of star formation in massive galaxies proceeds at a lower rate in higher mass galaxies and occurs almost instantaneously in lower mass galaxies. We believe that the last two of these scenarios are most plausible.

A future study will utilise a larger sample size to investigate in more detail the role of galaxy mass on environmental quenching, and incorporate a more detailed analysis of the galaxy group properties that drive the various quenching processes.

\section{Acknowledgements}
We would like to thank the anonymous referee for their constructive comments and suggestions, which improved the clarity and presentation of our results.

The SAMI Galaxy Survey is based on observation made at the Anglo-Australian Telescope. The Sydney-AAO Multi-object Integral field spectrograph (SAMI) was developed jointly by the University of Sydney and the Australian Astronomical Observatory. The SAMI input catalogue is based on data taken from the Sloan Digital Sky Survey, the GAMA Survey and the VST ATLAS Survey. The SAMI Galaxy Survey is funded by the Australian Research Council Centre of Excellence for All-sky Astrophysics (CAASTRO), through project number CE110001020, and other participating institutions. The SAMI Galaxy Survey website is \url{http://sami-survey.org/}.

The ARC Centre of Excellence for All-sky Astrophysics (CAASTRO) is a collaboration between The University of Sydney, The Australian National University, The University of Melbourne, Swinburne University of Technology, The University of Queensland, The University of Western Australia and Curtin University, the latter two participating together as the International Centre for Radio Astronomy Research (ICRAR). CAASTRO is funded under the Australian Research Council (ARC) Centre of Excellence program, with additional funding from the seven participating universities and from the NSW State Government's Science Leveraging Fund.

GAMA is a joint European-Australasian project based around a spectroscopic campaign using the Anglo-Australian Telescope. The GAMA website is \url{http://www.gama-survey.org/}.

ALS acknowledges support from a European Research Council grant (DEGAS-259586). 
SMC acknowledges the support of an Australian Research Council Future Fellowship (FT100100457).
SB acknowledges funding support from the Australian Research Council through a Future Fellowship (FT140101166)
JTA acknowledges the award of a SIEF John Stocker Fellowship.
MLPG. acknowledges support from a European Research Council grant (DEGAS-259586) and the Science and Technology Facilities Council (ST/L00075X/1).
MSO acknowledges the funding support from the Australian Research Council through a Future Fellowship Fellowship (FT140100255).
MA is funded by an appointment to the NASA Postdoctoral Program at Ames Research Centre, administered by Universities Space Research Association through a contract with NASA.
JvdS is funded under Bland-Hawthorn's ARC Laureate Fellowship (FL140100278).
NS acknowledges support of a University of Sydney Postdoctoral Research Fellowship.
This research made use of Astropy, a community-developed core Python package for Astronomy \citep{Astropy}. We also used the Numpy and Scipy scientific python libraries.

\end{document}